\documentclass[onecolumn,authoryear]{els-mrw} 

\usepackage{amsmath,amssymb,amsfonts,amsthm,makeidx,graphicx}
\usepackage{txfonts}
\usepackage{helvet}

\begin{document}

\chapter{Cosmology with Peculiar Velocity Surveys}\label{chap1}

\author[1]{Ryan J. Turner}%

\address[1]{\orgname{Swinburne University of Technology}, \orgdiv{Centre for Astrophysics and Supercomputing}, \orgaddress{John St.}}

\articletag{Chapter Article tagline: update of previous edition,, reprint..}

\maketitle

\begin{glossary}[Glossary]
\begin{tabular}{@{}lp{34pc}@{}}
\term{FP} &Fundamental Plane. A method of measuring distances to elliptical galaxies from their intrinsic properties.\\
\term{GR} &General Relativity. Theory that governs the gravitational interactions of all bodies on all scales. Can be tested by precisely measuring observables predicted by the theory.\\
\term{Gravitational Instability} &The theory by which slight inhomogeneities in the matter density field coalesce following the laws of gravity in order to form structure.\\
\term{$\Lambda$CDM} & Lambda Cold Dark Matter. The standard cosmological model, named for its primary constituents: $\Lambda$ -- Dark Energy, CDM -- Cold Dark Matter\\
\term{PV} &Peculiar Velocity. The motions of galaxies induced by gravitational interactions with their local environment, separate to their velocities due to the smooth expansion of the universe.\\
\term{RSD} &Redshift-space Distortions. Anisotropies in the clustering signal of galaxies as measured in redshift-space, caused by peculiar motion.\\
\term{SNe Ia} &Type Ia Supernovae. A specific form of supernova that occurs in binary star systems.\\
\term{TF} &Tully-Fisher. A method of measuring distances to spiral galaxies from their intrinsic properties.\\
\term{Velocity Correlation Function} &The Fourier transform of the velocity power spectrum, encoding information about the likelihood of galaxies to have coherent motion in some direction as a function of the separation between them.\\
\term{Velocity Power Spectrum} &An analogue of the traditional power spectrum, which instead captures information about fluctuations in the velocity field as a function of scale.\\
\term{ZOA} &Zone of Avoidance. An area of the sky that is obscured by the plane of the Milky Way, and thus hard to measure with galaxy surveys.\\

\end{tabular}
\end{glossary}

\begin{abstract}[Abstract]
Peculiar velocities are the motions of galaxies due to the gravitational influence of large-scale structure, and thus are an important cosmological probe of the underlying matter density field. In recent years the number of surveys designed to measure peculiar velocities has increased, to the point that it is plausible that we will have completely mapped out the peculiar velocity field in the local universe within the next decade. Such an abundance of data will enable us to place precise constraints on the growth rate of large-scale structure which in turn will inform us about the true nature of the laws of gravity and the standard cosmological model.

In this chapter, the physics governing the generation of peculiar velocities, the methods of measuring them, and the statistical tools used to extract cosmological information from them are described. It will also cover a swathe of current and future surveys dedicated to collecting peculiar velocities, what their aims are, and what these datasets may mean for the future of cosmological analyses.
\end{abstract}

\begin{BoxTypeA}[]{Key points}
\begin{itemize}
    \item The redshifts that we observe when measuring galaxies are not the `true' redshifts that we would expect for galaxies in a universe that is smoothly expanding according to the Hubble-Lema\^itre Law. The measured redshift is affected by a Doppler shift, caused by the peculiar velocities of galaxies. These velocities are induced by the gravitational influence of the environment on individual galaxies.
    \item The nature of these peculiar velocities means that, if we are able to measure them reliably, we can learn about the effects of gravity on the formation of large-scale structure in the universe, and potentially test the standard cosmological model itself.
    \item Peculiar velocities can be measured statistically through a phenomenon called redshift-space distortions, or directly if we can obtain redshift-independent estimates of the distances to nearby galaxies. There are several ways of doing this, but the most common involve Type Ia supernovae or galaxy scaling relations, such as the Tully-Fisher relation or Fundamental Plane relation.
    \item Teams of astronomers have worked together to produce large catalogues of these direct peculiar velocity measurements, in what we call `Peculiar Velocity Surveys'. The data in these catalogues, when used in tandem with more typical galaxy redshift surveys, can produce stringent constraints on a cosmological parameters such as the growth rate of large-scale structure. The cosmological information captured by such surveys can be extracted by utilising our statistical knowledge about the measured density and velocity fields, and the relationship between the two fields. These include the two-point correlation function, power spectra, and linear reconstructions of the velocity field.
    \item In the near future, it is expected that we will have measured hundreds of thousands of galaxy peculiar velocities in the local universe. Forecasts predict that such a vast amount of data will provide us with measurements of the growth rate that are precise enough to rule out alternative theories of gravity and extensions to the standard cosmological model.
\end{itemize}
\end{BoxTypeA}
\section{Introduction}\label{sec:intro}
When astronomers measure the redshifts of galaxies, using photometric or spectroscopic techniques, they are not measuring the `true' distances to those galaxies. Instead, they are measuring a combination of redshifts caused by two different phenomena:
\begin{itemize}
    \item The recessional velocity of the galaxy, approximated as $c z_{\mathrm{cos}}$ at low redshift, due to the smooth expansion of the universe as predicted by the Hubble-Lema\^itre Law
    \item The peculiar velocity of the galaxy, $v_{\rm{pec}}$, due to the gravitational influence of the local environment on the galaxy.
\end{itemize}
Thus, the observed redshift of an object $z_{\rm{obs}}$ is not its cosmological redshift $z_{\rm{cos}}$, but a Doppler-shifted value modified by its peculiar motion
\begin{equation}\label{eq:PV}
    (1 + z_{\rm{obs}}) = (1 + z_{\rm{cos}})(1 + v_{\rm{pec}}/c)
\end{equation}
where $c$ is the speed of light \citep[see][for a more thorough understanding of this equation]{Davis2014}.

While this may seem like an annoyance, the method in which peculiar velocities (peculiar here meaning specific to the individual galaxies, and not `strange') are imparted onto galaxies means that they hold a wealth of cosmological information pertaining to gravitational physics and the underlying matter density field of the universe. 
The potential information held that may be unlocked by a deeper understanding of these velocities has led astronomers in recent years to undertake a series of surveys designed to measure as many galaxy peculiar velocities as is feasible to do in the local universe. Referring to Eq. \ref{eq:PV}, if we can obtain a redshift-independent measurement of the distance to a galaxy to use as a substitute for $z_{\rm{cos}}$ then we can disentangle the effects of cosmological expansion and peculiar motion on the observed redshift to directly measure the peculiar velocity of the galaxy.
There are many ways of doing this -- utilising scaling relations between the properties of different groups of galaxies, using the intrinsic brightness of a specific classification of supernovae, and using the signals of gravitational waves caused by the merging of distant compact objects, to name only a few. As well as measuring peculiar velocities directly, we can also measure them statistically from redshift-space distortions in the clustering statistics of galaxies. The positions of galaxies in redshift-space are inferred from their observed redshift, which includes the radial $v_{\rm{pec}}$ component. This introduces anisotropies in the clustering pattern along the line of sight on small scales -- the `Finger-of-God effect' -- and on large scales -- the `Kaiser effect' \citep{Kaiser1987}.

Contextualising these velocities as part of the local velocity field and using the relationship of this field to the matter density field -- employing the two-point statistics of both fields and making direct comparisons between observations in one field against inferences made about that field based on observations of the other -- astronomers can begin to understand the role of peculiar velocities in the evolution of the universe. Measuring peculiar velocities for a large sample of galaxies over a vast area of the sky therefore gives astronomers an extremely powerful tool for understanding how large-scale structure in the universe has formed over cosmic time, for measuring cosmological parameters such as the matter density $\Omega_m$, and potentially shedding light onto the true nature of dark energy.

In this chapter we will explore the growing field of peculiar velocity cosmology, the current and future surveys dedicated to obtaining peculiar velocities for as many galaxies as possible, and how astronomers use that data to understand the universe. The mathematics describing the relation between peculiar velocities and the density field, and how this relates to the growth rate of large-scale structure, are summarised in Section \ref{sec:PVs}. In Section \ref{sec:measures} we cover the various methods used to obtain measurements of peculiar velocities, including galaxy scaling relations and standard candles, while in Section \ref{sec:surveys} we summarise some of the recent major peculiar velocity surveys that utilise these methods. In Section \ref{sec:cosmo} we cover some of the methods that employ peculiar velocity data which astronomers use to measure the growth rate of large-scale structure and other cosmological parameters, constrain theories of modified gravity, and improve estimates of the Hubble constant. We give an overview of the current state of the field, including results from the aforementioned methodologies, and a view to the future of the field of peculiar velocity cosmology in Section \ref{sec:current}, before finally summarising in Section \ref{sec:summary}.

\section{What are Peculiar Velocities?}\label{sec:PVs}
In the first moments of the universe, all of the matter and energy contained within it were in an initial state of exceedingly high temperature, density and pressure.
The universe then underwent a process of rapid expansion, and thus cooling, which then slowed but crucially did not stop. We refer to this process as the Big Bang model of the formation of the universe, and we collect the different components of energy and matter that make up the universe in our standard (or `concordance') cosmological model $\Lambda$CDM. In this understanding of the universe, the laws of gravity are governed by Einstein's theory of general relativity (GR) and structures such as galaxies and galaxy clusters form according to the theory of gravitational instability.

One important element of the Big Bang model that must be included to match our observations of the primordial universe is the inflationary period. Inflation, a period of accelerated, exponential, expansion that the universe experienced around 10$^{-35}$ seconds after the Big Bang, was proposed to resolve the horizon and flatness problems that had been the two largest issues with the Big Bang model.

Inflation occurred quickly, lasting for all of $10^{-33}$ seconds, but had massive effects on the size of the universe. By the time inflation had subsided, fractions of a second after of the Big Bang, the size of the universe had increased by approximately 60 `e-folds' or by a factor of around 10$^{27}$. This is equivalent to a water molecule growing to the size of Omega Centauri in roughly one-trillionth of the time it takes light to cross a hydrogen molecule.

For our purposes, the key result from this inflationary period is how it set the scene for the large-scale structure we see in the universe today. The universe prior to inflation was microscopic, and any fluctuations in the temperature/density of the universe were quantum in nature. During inflation, these quantum fluctuations grew with the rest of the universe to become the initial fluctuations in the density field that existed on astronomical scales. These primordial density fluctuations are the seeds from which density perturbations and eventually large-scale structure in the universe formed.

We can then relate this formation of structure to the peculiar motions of galaxies within the theory of gravitational instability. Gravitational instability states even the smallest perturbation in the matter density field will grow under the influence of gravity. Overdense regions in the density field will exert a greater gravitational force on its surroundings than less dense regions, and so more matter is drawn towards them further increasing their density. The expansion of the universe, as caused by dark energy, has the opposite effect, and acts to inhibit the growth of structure. If the universe were to be expanding rapidly there is less time for matter contained within it to interact under gravity and coalesce, and vice versa. The growth of structures we see today like stars, galaxies, clusters, and filaments, formed over cosmic timescales, is the result of this balance of gravity and expansion.

We can understand this mathematically, starting with the definition of overdensity at some point in space $\textbf{x}$ and time $t$:
\begin{equation}\label{eq:contrast}
    \delta(\textbf{x},t) \equiv \frac{\rho(\textbf{x},t) - \bar{\rho}(t)}{\bar{\rho}(t)}
\end{equation}
where $\rho(\textbf{x},t)$ is the mass density and $\bar{\rho}(t)$ is the mean density of the universe. By treating the density field as a perfect fluid, describable only by its energy density $\rho$ and pressure $p$, we can dictate its evolution using the three fluid equations. 

The Continuity equation:
\begin{equation}
    \frac{\partial\rho}{\partial t} + \nabla\cdot(\,\rho\textbf{v}\,) = 0\,,
\end{equation}
where \textbf{v} is velocity. The Continuity equation ensures the conservation of mass and states that the growth of density in a volume is equivalent to the amount of matter entering the volume. If mass flows into (out of) a region, there is positive (negative) divergence and the density increases (decreases). This equation stands for non-relativistic matter, in the relativistic case we must substitute $\rho$ for $\left(\rho + P/c^2\right)$ in the second term.

The Euler equation:
\begin{equation}
    \frac{\partial \textbf{v}}{\partial t} + (\,\textbf{v}\cdot\nabla)\textbf{v} + \frac{1}{\rho}\nabla P + \nabla\Phi = 0\,,
\end{equation}
where $\Phi$ is the gravitational potential. The Euler equation states that gravitational forces and forces due to pressure are what induce velocity flows in the universe, and can be thought of as an application of Newton's second law. 

The Poisson equation:
\begin{equation}
    \nabla^2\Phi = 4\pi G \rho,
\end{equation}
where $G$ is the gravitational constant. The Poisson equation describes the source of the gravitational potential $\Phi$ as being induced by the matter and energy content of the universe. Regions with higher matter density will produce stronger gravitational forces, and vice versa.

From these three equations it is possible for us to arrive at the general equation for the evolution of density fluctuations:
\begin{equation}\label{eq:gen_densfluc}
    \frac{\partial^2\delta}{\partial t^2} + 2\frac{\dot{a}}{a}\frac{\partial\delta}{\partial t} = \frac{\nabla^2p}{\rho a^2} + \frac{1}{a^2}\nabla\cdot(1 + \delta)\nabla\phi\,.
\end{equation}
In Eq. \ref{eq:gen_densfluc} we have introduced $a(t)$, which is known as the scale factor. This function describes how distances between objects in the universe change as the universe expands, and by convention the value of the scale factor today is defined to be unity, $a(t_0) = 1$, and is zero at the time of the Big Bang. Following this convention the relation between the scale factor and redshift is given by $a = 1/(1+z)$. We have implicitly included the time dependence of $a(t)$ in Eq. \ref{eq:gen_densfluc}, and have used the convention that an overdot above a value indicates that the value is taken as a derivative with respect to time. That is, $\dot{a} = da/dt$.

If $\delta << 1$, and assuming that the velocity field is irrotational (contains no vorticity), we can also define the linearised density perturbation equations:
\begin{equation}\label{eq:linpert1}
    \frac{\partial^2\delta}{\partial t^2} = 2\frac{\dot{a}}{a}\frac{\partial\delta}{\partial t} - \frac{\nabla^2p}{\rho a^2} - 4\pi G\rho\delta\,,
\end{equation}
\begin{equation}\label{eq:linpert2}
    \frac{\partial \delta}{\partial t} = -\frac{1}{a}\nabla\cdot\textbf{v}\,.
\end{equation}
Equation \ref{eq:linpert1} is a second order differential equation, and so the most general solution for $\delta$ is given by:
\begin{equation}
    \delta(\textbf{x},t) = \delta(\textbf{x})D_1(t) + \delta(\textbf{x})D_2(t)\,,
\end{equation}
where $D_1$ is a growing mode and $D_2$ is a decaying mode. Measured today, the decaying solution is wholly subdominant and thus the solution to the perturbation equation is entirely reliant on the growing mode $D_1$. Substituting this result for $\delta$ into Eq. \ref{eq:linpert2}, we find:
\begin{equation}\label{eq:deltaD1}
    \frac{\partial \delta}{\partial t} = \delta(\textbf{x})\frac{dD_1}{dt} = -\frac{1}{a}\nabla\cdot\textbf{v}\,.
\end{equation}

The universe has spent the majority of its existence in the `matter-dominated' era, from matter-radiation equality until dark energy became dominant around 10 billion years later, and it was during this time that most of the structure in the universe formed. In the matter-dominated era, $a(t) \propto t^{2/3}$ and the Hubble parameter $H(t) \equiv \dot{a}(t)/a(t) \propto 2/(3t)$, and by assuming that $\delta(t) \propto t^n$ we can rewrite the growth equation as
\begin{equation}
    n(n - 1) + \frac{4}{3}n - \frac{2}{3} = 0
\end{equation}
which has two solutions: $n = 2/3$ and $n = -1$. What we've found is that the growing mode solution during the matter-dominated era grows as $t^{2/3}$ (i.e. analogously to the scale factor during this period) and the decaying solution tails off as $t^{-1}$, in line with our assumptions that lead to Eq. \ref{eq:deltaD1}.

The velocity associated with the growing mode is:
\begin{equation}\label{eq:vtof}
    \textbf{v} = \frac{Hf\textbf{g}}{4\pi G\rho} = \frac{2f\textbf{g}}{3H\Omega}
\end{equation}
where $\textbf{g}$ is the peculiar gravitational acceleration, i.e. the gradient of the potential perturbation $\phi$, given by $\phi(\textbf{x},t) = \Phi(\textbf{r},t) - (1/2)a\ddot{a}x^2$. $\Omega = 8\pi G\rho/(3H^2)$ is the dimensionless density parameter, and $f$ is a parameter called the linear growth rate of structure. We define $f$ as
\begin{equation}
    f \equiv \frac{d\ln D_1(a)}{d\ln a} = \frac{1}{D_1(a)}\frac{d}{d\ln(a)}D_1(a)\,,
\end{equation}
and with this convention we can rewrite Eq. \ref{eq:deltaD1} as
\begin{equation}\label{eq:divofv}
    \nabla\cdot\textbf{v} = -aHf\delta(\textbf{x},t)\,.
\end{equation}
Integrate over all positions, and we arrive at an equation for the peculiar velocity:
\begin{equation}\label{eq:pv-grav}
    v(\textbf{r}) = \frac{H_0af}{4\pi}\int d^3\textbf{r}'\frac{\delta(\textbf{r}')(\textbf{r}' - \textbf{r})}{|\textbf{r}' - \textbf{r}|^3}
\end{equation}

All of this is to say that the peculiar velocity field $v(\textbf{r})$ measured at some position $\textbf{r}$ is due to the gravitational influence of the surrounding density field $\delta(\textbf{r}')$ measured at all positions $\textbf{r}'$. The more distant any considered point in the density field $\textbf{r}'$ is from $\textbf{r}$, the less effect it will have on $v(\textbf{r})$ as dictated by the $(\textbf{r}' - \textbf{r})/|\textbf{r}' - \textbf{r}|^3$ term. The growth rate of structure, $f$, as the name would suggest, dictates the rate at which structure in the universe is able to form. If $f$ were to be large, density perturbations grow more quickly, increasing the strength of gravitational attraction in these overdense regions and in turn the peculiar velocities that they induce. If instead $f$ were small, the opposite case would be true resulting in peculiar velocities that are smaller.

We can also understand peculiar velocity in terms of the relationship between physical and co-moving coordinates. We can relate the physical coordinates of a point in space-time $\textbf{r}(t)$ to the co-moving coordinates $\textbf{x}(t)$ via
\begin{equation}
    \textbf{r} = a(t)\textbf{x}\,.
\end{equation}
We can then determine the proper velocity as:
\begin{equation}\label{eq:propv}
    \textbf{u} = \frac{d\textbf{r}}{dt} = \dot{a}\textbf{x} + a\textbf{v}(\textbf{x},t) = \left(\frac{\dot{a}}{a}\right)\textbf{r} + a\textbf{v}(\textbf{r}/a, t)\,,
\end{equation}
where $a\textbf{v}$ is the peculiar velocity, the change in the co-moving position \textbf{x}. Eq. \ref{eq:propv} is reminiscent of the Hubble-Lema\^itre law but with an additional term included. The first term, ($\dot{a}/a)\textbf{r} = H\textbf{r}$, is attributed to smooth expansion of the universe. The second term which causes the proper velocity to deviate from what we would expect from the smooth Hubble expansion alone is due to the peculiar velocity, which we have established is caused by the gravitational influence of local overdensities. This is borne out in the data, and almost all galaxies will deviate from the Hubble-Lema\^itre law to some extent.

\subsection{The growth rate of large scale structure}
Peculiar velocities are seen as a key cosmological probe for their sensitivity to, and ability to constrain, the growth rate of structure $f$. We will cover this briefly here and refer readers to \cite{Huterer2023} and references therein for a more thorough review of this aspect of the field. We can immediately see this from Eq. \ref{eq:pv-grav}. If we compare estimates of the mass density field $\delta(\textbf{r})$ obtained from galaxy redshift surveys and the peculiar velocity field $v(\textbf{r})$ obtained from peculiar velocity surveys, then we can constrain $f$. As the growth rate represents how matter coalesces under the influence of gravity, it can also be used as evidence for or against GR and therefore the standard cosmological model. One typical parameterisation of the growth rate is as a function of the mass density parameter $\Omega_m$:
\begin{equation}\label{eq:grindex}
    f(z) = \Omega_m(z)^{\gamma}\,,
\end{equation}
where $\gamma$ is the growth index, whose value is determined by the considered theory of gravity. For example, in the standard concordance model of cosmology $\gamma_{\Lambda\mathrm{CDM}} = 6/11$, while in the `Dvali-Gabadadze-Porrati' (DGP) model of gravity $\gamma_{\mathrm{DGP}} = 11/16$. It should also be noted that Eq. \ref{eq:grindex} is a close approximation to the exact solution of Eq. \ref{eq:linpert1}. One issue with this concept that we are yet to resolve is that the matter density field in Eq. \ref{eq:pv-grav} describes contributions from all matter, including the dark matter that we cannot detect. We can, however, measure the density of galaxies $\delta_g$ from observations. We know that galaxies and galaxy clusters reside in large halos of dark matter, evidenced by the kinematics of the galaxies themselves and their motions within these clusters. If we assume that distribution of galaxies linearly traces the underlying distribution of dark matter, we can describe the density fluctuations in the number of galaxies we observe as:
\begin{equation}\label{eq:bias}
    \delta_g(\textbf{r}) = b\,\delta(\textbf{r})
\end{equation}
where $\delta$ is the matter density shown in Eq. \ref{eq:pv-grav} and $b$ is the linear galaxy bias parameter. Replacing the density in Eq. \ref{eq:pv-grav} with the galaxy density $\delta_g$, we no longer measure $f$ but instead a combination of the growth rate and the linear galaxy bias $\beta(z) = f(z)/b$, called the linear redshift distortion parameter. We do not always know the value of $b$ a priori, therefore we often reparameterise our results in terms of $\sigma_8$, the root mean square of ($z=0$) density fluctuations on $8$ h$^{-1}$ Mpc scales, where
\begin{equation}\label{eq:sigma8}
    \sigma_8^2 = \frac{1}{2\pi^2}\int dk\,k^2\,P(k)\,|W(kR_8)|^2
\end{equation}
where \textit{P(k)} is the matter power spectrum measured at redshift z = 0, and $W(kR_8)$ is the Fourier transform of the spherical top hat window function of radius $R_8 = 8$ h$^{-1}$ Mpc. We can then equate
\begin{equation}
    f\sigma_8 = \beta\sigma_8^g\,,
\end{equation}
where $\sigma_8^g = b\sigma_8$ is the rms fluctuation in the galaxy overdensity field within spheres of $8$ h$^{-1}$ Mpc, and can be measured directly from galaxy redshift surveys.

\begin{figure}
    \centering
    \includegraphics[width=0.75\linewidth]{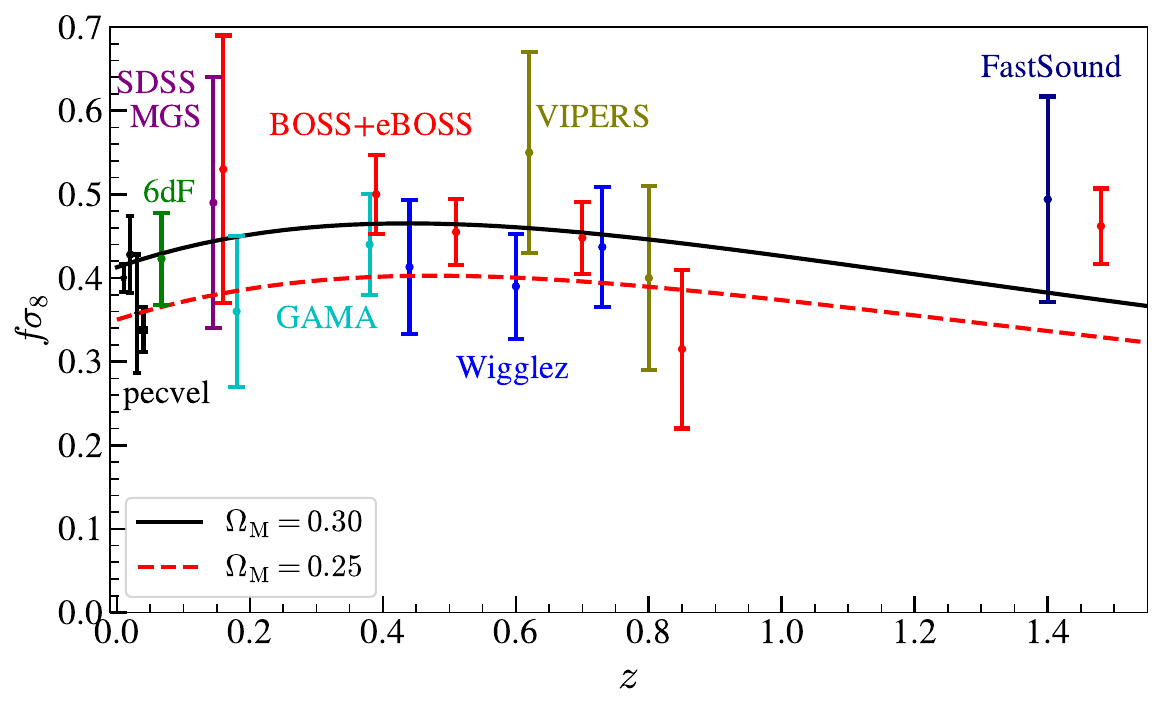}
    \caption{A collection of $f\sigma_8$ measurements made by different collaborations and groups, using RSD and PV datasets, shown as a function of redshift. Error bars are similarly coloured to their corresponding surveys. Measurements made using peculiar velocity data are labelled `pecvel' and are shown in black. Also plotted are lines showing $\Lambda$CDM predictions for $f\sigma_8$ for different values of $\Omega_M$, the solid line showing $\Omega_M = 0.30$ and the dotted line showing $\Omega_M = 0.25$. Credit: Huterer (2023). }
    \label{fig:fs8}
\end{figure}

The combination of parameters $f$ and $\sigma_8$ is commonly referred to as the normalised growth rate of large scale structure $f\sigma_8$, and the two are considered degenerate parameters. We can understand this intuitively by imagining the effects that different values of $f$ and $\sigma_8$ may have on the peculiar velocity field. We've already discussed how larger values of $f$ imply that density perturbations grow more quickly, resulting in more overdense regions, but this can also be explained by the rms fluctuation in the density field being larger. In essence, for any given value of $f\sigma_8$ at a fixed redshift, a higher value of $f$ can be compensated by a smaller value of $\sigma_8$, or vice versa, with no meaningful way to disentangle the two effects other than to make measurements spanning a range of redshifts. In Fig. \ref{fig:fs8} we show recent measurements of $f\sigma_8$ made by several large collaborations and other groups, plotted at the effective redshifts of their datasets, and how those measurements can be used to compare against $\Lambda$CDM cosmology for given values of the matter density $\Omega_m$.

The efficacy of peculiar velocities as a cosmological probe is dictated by the accuracy of our distance measurements and the number of tracer galaxies available. The cumulative total of supernovae distances is small in comparison to the total number of distances measured with galaxy scaling relations, and these relations are limited to low redshifts ($z \lesssim 0.15$) by the signal-to-noise ratio of the data. These distance measures are discussed further in Section \ref{sec:measures}. 

\begin{figure}
    \centering
    \includegraphics[width=0.75\linewidth]{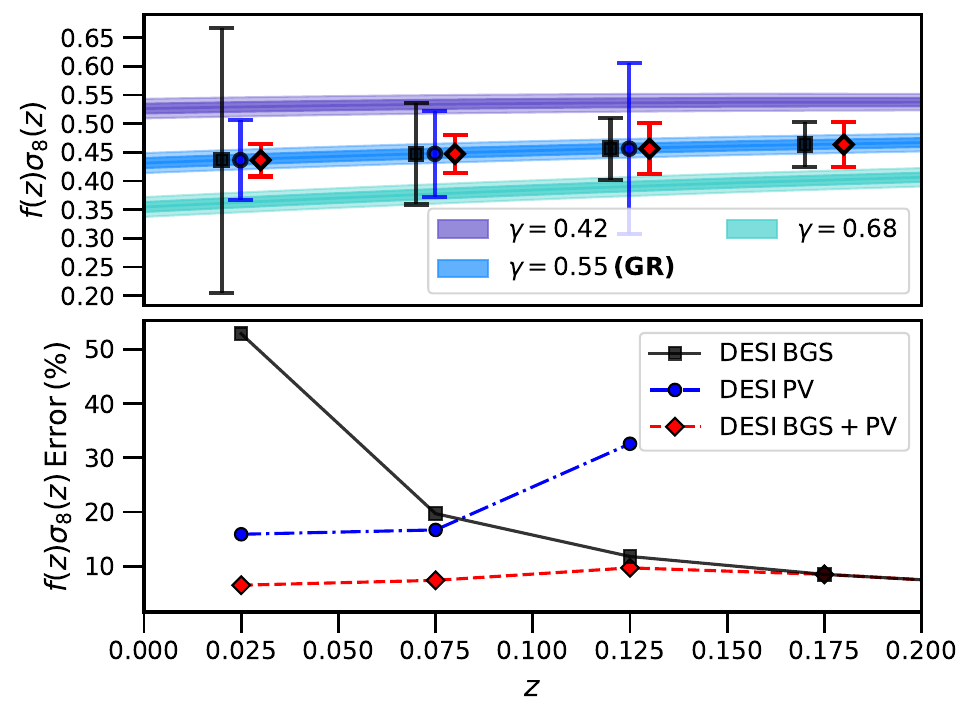}
    \caption{Forecasts for constraints on the combined parameter $f\sigma_8$ in different redshift bins, $\Delta z = 0.05$, using different DESI datasets. In black are constraints made using only the DESI Bright Galaxy Survey (BGS), only the DESI Peculiar Velocity (PV) survey in blue, and the combined BGS + PV two-field constraints in red. The forecasts in the top panel are slightly offset from the bin midpoints for visibility, and are centred on the GR prediction for the growth index $\gamma = 0.55$. Other coloured bands show different parameterisations. The bottom panel shows the relative error in each forecast, in each redshift bin. Credit: Saulder et al. (2023).}
    \label{fig:desi-forecast}
\end{figure}

The entirety of peculiar velocity cosmology happens in this small cosmological volume, which means that we are sensitive to fewer of the large-scale wavelengths that constitute the density field in Fourier space, otherwise known as Fourier modes. This introduces an amount of sample variance to our statistics, which imposes a fundamental lower bound on the uncertainty of our measurements. This is true for measurements made purely from the velocity field, however we can get around this when combining the velocity and density fields in a `two-field' analysis. Relating the two fields via gravitational instability to instead constrain $\beta$, we instead constrain the ratio of two different tracers which is independent of the underlying matter distribution. This significantly reduces the contribution of sample variance to the error budget, while not removing it entirely because of measurement noise.
This is the key characteristic of peculiar velocity surveys that make them competitive with other probes of cosmic growth. Galaxy redshift surveys that are used to measure redshift-space distortions are inherently limited by sample variance in the large-scale modes at $z = 0$, as they cannot employ the `multi-tracer' approach which is available to peculiar velocity analyses and thus have no means to remove this error term. The two-field, multi-tracer approach is the most promising means by which we can constrain the growth rate, with measurement accuracy forecasted to be similar to that obtained for measurements of cosmic expansion, i.e. of order $\lesssim 3\%$ \citep{Koda2014, Howlett2017}. Fig. \ref{fig:desi-forecast} shows such a set of forecasts made for the Dark Energy Spectroscopic Instrument (DESI) Bright Galaxy Survey and Peculiar Velocity survey. It is clear from these forecasts that constraints from the combined-tracer approach is markedly improved in comparison to those from the Bright Galaxy Survey alone. Producing measurements as accurate as this will allow for significantly better distinction between different $\gamma$-parameterisations of gravity, following Eq. \ref{eq:grindex}, at low redshifts.

\subsection{Why are peculiar velocities useful?}
While the obvious use case for peculiar velocities is in constraining the growth rate they are a useful tool in other regards too, such as for testing gravitational physics, charting the flow of velocities in the local universe, and measuring the Hubble constant. 

The velocity field in Fourier space is related to the matter overdensity field in Fourier space via the equation:
\begin{equation}\label{eq:fourierv}
    \Tilde{\textbf{v}}(\textbf{k}) = -iaHf\frac{\textbf{k}}{k^2}\Tilde{\delta}_m(\textbf{k})\,,
\end{equation}
which we can simply understand as $\Tilde{\textbf{v}}(\textbf{k}) \propto \Tilde{\delta}_m(\textbf{k})/k^2$. This equation is equivalent to Eq. \ref{eq:divofv}, rather than being a separate relation between velocity and density. The factor of $i$ appears because the velocity field is the gradient of the potential, as shown in Eq. \ref{eq:vtof}, and the gradient operator appears as a factor of $i\textbf{k}$ in Fourier space. Physically, this manifests as a phase shift between the velocity field and the matter overdensity field, due to the fact that the velocity is derived from gradients in the field rather than the field itself. The velocity field contains two fewer factors of the wavenumber $k$, and thus is more sensitive to smaller values of $k$, and thus larger scales with respect to the density field. The density field, meanwhile, contains more power at smaller scales. This characteristic of the velocity field means that it is sensitive to large-scale modes that the density field is incapable of accessing. All of this is to say that the signal is greater at differing values of $k$ for each field, and taken in combination can reduce sample variance in our measurements. This sensitivity to larger scale modes is what makes the peculiar velocity field a powerful probe of the underlying matter density field, theories of modified gravity, and alternative models of dark energy that is complementary to other techniques such as redshift-space distortions that can accurately measure growth on intermediate scales. 

Theorists have been encouraged to develop alternative models to General Relativity (GR) based on the existence of dark energy. This model requires a dark sector that we cannot directly observe and accounts for approximately $95\%$ of the total mass-energy budget, around $70\%$ of that total being in the form of dark energy. Dark energy itself is purely phenomenological, meaning that we have chosen to represent it with $\Lambda$, the cosmological constant, with no physical justification or reasoning beyond the fact that it fits our observations. This uncertainty at the heart of the standard cosmological model is what drives the creation of modified theories of gravity, or alternatives to GR, to either provide an explanation for dark energy or remove the need for the dark sector entirely. 

In addition to being parameterised via the growth index as shown in Eq. \ref{eq:grindex}, another potential indication of additional physics may lie in the behaviour of the growth rate as a function of scale. This particular dependence on scale cannot be modelled by $\gamma$. The specifics of modified gravity models are beyond the scope of this chapter, we direct readers to \cite{Jain2010} or \cite{Joyce2016} for a more complete overview of these models, but to summarise the key aspects: in particular models of modified gravity, such as $f(R)$ gravity which modifies the Einstein-Hilbert action present in GR, the gravitational constant $G$ is replaced with an `effective' gravitational constant $G_{\mathrm{eff}}$ which is a scale-dependent function. This scale-dependence in $G_{\mathrm{eff}}$ propagates into a scale-dependence in the growth of density perturbations and hence the growth rate. Thus, measuring the growth rate accurately across multiple scales is an additional test of gravity.

One constraint that these theories must abide by is that there is a region of space where we know the laws of gravity, as predicted by GR, work extremely well -- the solar system. Any new theories must exert their differences on larger-than-solar-system scales but simultaneously `screen' those changes once we get down to these smaller scales, so as to reproduce our observations. `Screening mechanisms', as they are called, mean that any evidence of deviations from the standard model and GR can only be observed on very large scales where the effects of any changes may actually manifest. Peculiar velocities, being excellent tracers of fluctuations in the matter density on large scales, are a promising tool to confirm or rule out the existence of new physics beyond GR.

\begin{figure}
    \centering
    \includegraphics[width=0.75\linewidth]{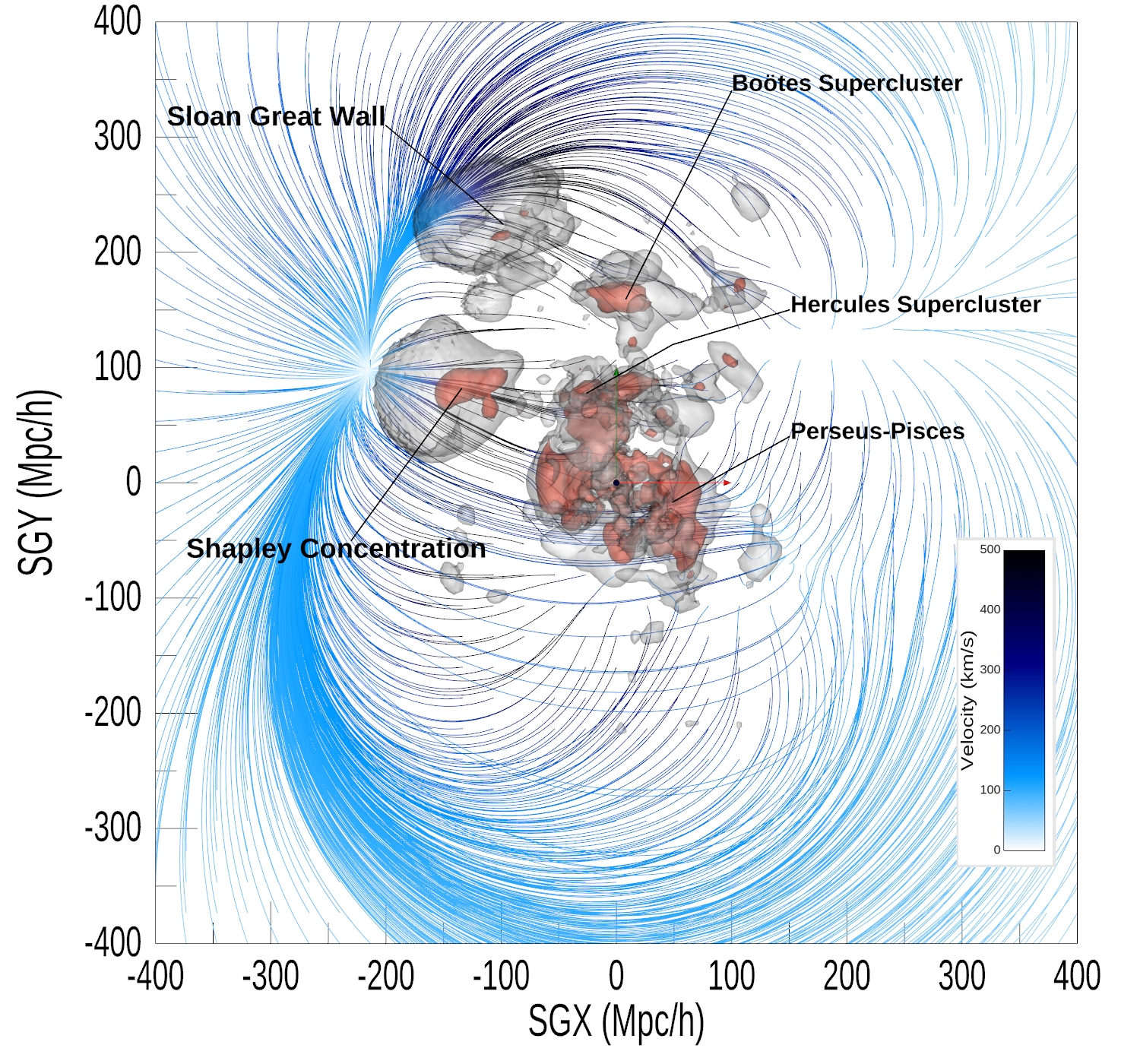}
    \caption{A reconstruction of the velocity field made using data from the Cosmicflows-4 catalogue. The blue `streamlines' represent the velocity field, different colours corresponding to different amplitudes in the field. Isosurfaces represent areas of differing overdensity, $\delta = 0.2$ in grey and $\delta = 0.5$ in red. Labels indicate noteworthy regions in the nearby universe. Credit: Hoffman et al. (2024).}
    \label{fig:cf4-velfield}
\end{figure}

Peculiar velocity measurements can also be used to map the `cosmography' of the local universe. By measuring the velocities of galaxies in some sample of nearby galaxies we can begin to plot a larger, overarching `map' of the movements of galaxies and how they are affected by large-scale structure. These plots end up being reminiscent of weather maps, where arrows and isobars can represent the direction and strength of wind, and enable us to chart out where regions of overdensity lie in the local universe. In Fig. \ref{fig:cf4-velfield} we show an example of such a plot made by \cite{Hoffman2024} using Cosmicflows-4 data, where the velocity field is represented by blue `streamlines' and overdensities are represented by grey and red regions. Using these maps we have been able to determine the location of the so-called `Great Attractor', a large area of space that exerts great gravitational influence that we now know to be obscured by the `Zone of Avoidance' (ZOA), a portion of the night sky hidden by the plane of the Milky Way. Other regions of high density, or superstructures, include the Coma supercluster, the Shapley supercluster, the Bo\"{o}tes supercluster, and the Sloan Great Wall.

While we can use distance indicators like supernovae to obtain distances and therefore peculiar velocities, we can flip this scenario and use our understanding of the local peculiar velocity field to infer an accurate estimate of the host galaxy redshift for these standard candles. Using the distances inferred from the standard candles themselves, but using a independent estimate of the peculiar velocity of the host galaxy to better measure its redshift, we can produce a more accurate distance-redshift diagram for a sample of standard candles and subsequently obtain tighter constraints on the Hubble constant, $H_0$. This is done for standard candles and standard sirens, where different models of the peculiar velocity field have been used to infer the cosmological redshift for hosts of Type Ia supernovae and for the host of the gravitational wave event GW170817 in order to measure $H_0$ more accurately. Peculiar velocities can introduce additional error to the Hubble diagram, and it is often the case that standard candles at sufficiently low redshift are excluded from $H_0$ determinations. Incorporating more sophisticated and complete maps of the peculiar velocity field into standard candle analyses is an active area of the field.

\section{How do we measure Peculiar Velocities?}\label{sec:measures}
As mentioned above, to measure the peculiar velocity of a galaxy we require a redshift-independent distance indicator so that we can separate the observed redshift into its cosmological and peculiar components. We can categorise these into 'Primary' and 'Secondary' distance indicators, based on how the methods are calibrated. 
Primary distance indicators are calibrated from intra-galactic distance measurements, using techniques such as parallax, and once calibrated can be used to infer distances. These indicators include Cepheid variable stars and Tip of Red Giant Branch stars.
Secondary distance indicators must be calibrated using primary distance indicators in order to set the zero-point for the method before they themselves can be used to infer distances. Examples of secondary indicators include Type Ia supernovae, and galaxy scaling relations such as the Tully-Fisher and Fundamental Plane relations. 

\subsection{Galaxy scaling relations}
Galaxy scaling relations are the most commonly used form of distance indicator, as we can more easily measure large samples of galaxies relative to supernovae or gravitational waves. These relations link distance independent qualities of galaxies, pertaining to some aspect of their internal velocity, to a distance dependent measurement like absolute magnitude or effective radius. We can broadly split galaxies into 'spirals' and 'ellipticals', and thankfully there exists a scaling relation for both galaxy types. 
\subsubsection{Tully-Fisher Relation}\label{ssec:TF}
\begin{figure}
    \centering
    \includegraphics[width=0.75\linewidth]{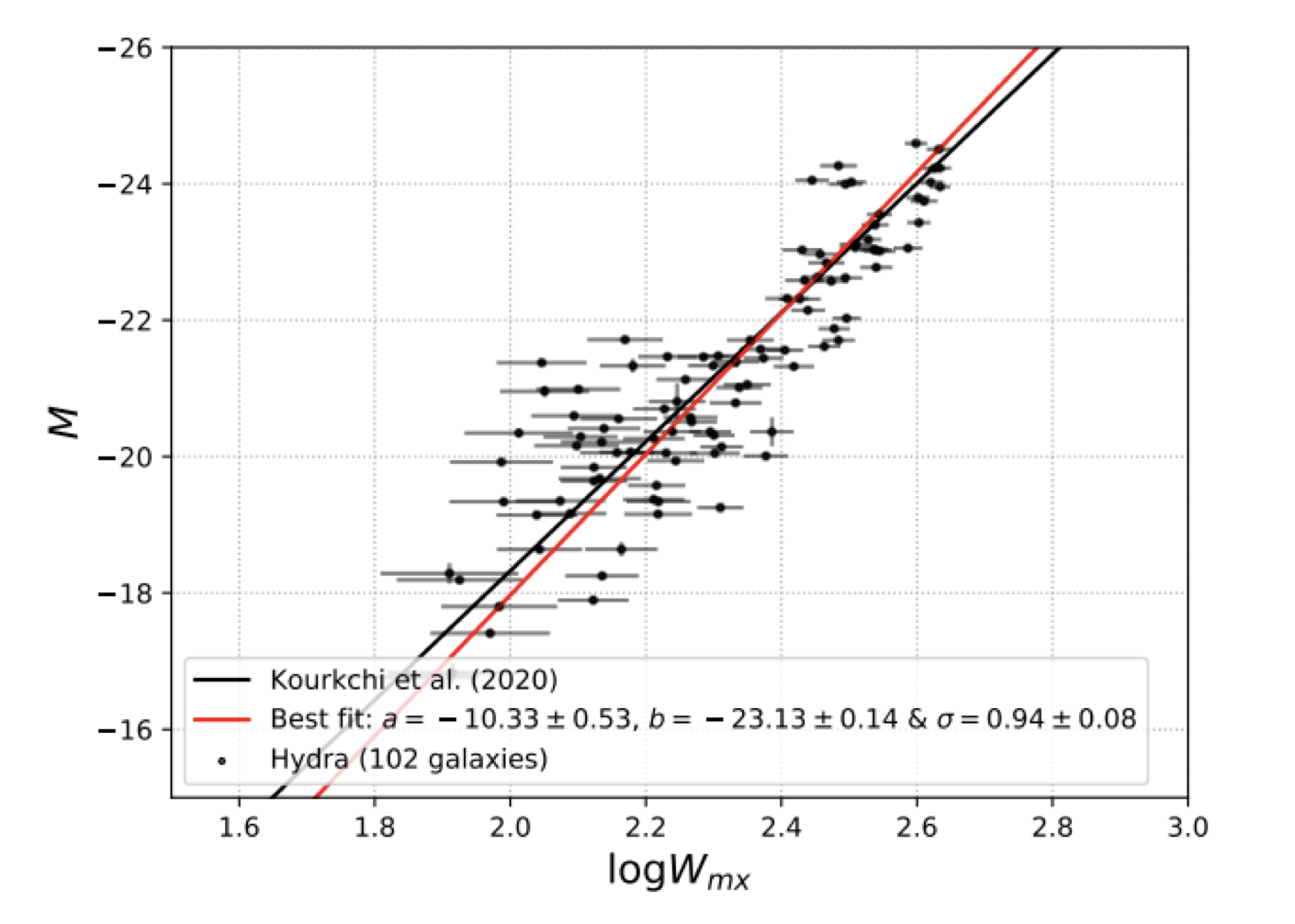}
    \caption{An example of the Tully-Fisher relation, measured for galaxies in the Hydra cluster using unWISE photmetry. The x-axis describes the logarithm of the linewidth of the observed HI data, where $W_{mx}$ is approximately equal to twice the maximum rotation velocity of the galaxy before accounting for its inclination angle. The black line is the calibrated Tully-Fisher relation from \cite{KourkchiTFR}. Courtois et al. (2023).}
    \label{fig:TFR}
\end{figure}

The Tully-Fisher (TF) relation \citep{Tully1977} links the rotational velocity of spiral galaxies to their total luminosity via the absolute magnitude of the galaxy. The measured luminosity of the galaxy via its apparent magnitude is a distance-dependent observable, and by comparing it to predictions of the galaxy's absolute magnitude we can determine a distance. The TF relation must be calibrated using an existing sample of galaxies with accurately known distances, measured using a primary distance indicator, and has the form
\begin{equation}
    M = a\,\log_{10}(V) + b\,,
\end{equation}
where $M$ is the absolute magnitude of the galaxy measured in some band and $V$ is the rotational velocity. See Fig. \ref{fig:TFR} for an example of a TF relation constructed from galaxies in the Hydra cluster \citep{CourtoisTF}. The coefficients $a$ and $b$ are the slope and zeropoint of the TF relation respectively, which can be fit using the calibration sample of galaxies. Once we have constructed the TF relation we can then use it to determine the properties of other galaxies. If we know the rotational velocity of a galaxy, e.g. from the width of the HI 21cm profile, we can predict its absolute magnitude, and then by comparing this prediction to the galaxy's apparent magnitude we can obtain a distance.
There is scatter inherent to the Tully-Fisher relation of around $0.35 - 0.40$ mag, this is equivalent to approximately $20\%$ uncertainty on the distances measured with the TF relation. 
While we do not completely understand the precise physical basis for the TF relation, we know that it is predicated on the notion that spiral galaxies are flattened, rotation-supported systems, thus the rotation velocity of these galaxies measured at some radius from their centre is related to the total mass contained within this radius. \cite{Strauss1995} presented a simple argument based on the assumption that galaxies are virialised systems, and their dynamics can be expressed by the proportionality
\begin{equation}
    v_{\mathrm{rot}}^2 \propto \frac{M}{R}\,.
\end{equation}
where $M$ is the mass within a sphere of radius $R$, $R$ is the distance from the centre of the galaxy, and $v_{\mathrm{rot}}$ is the rotational velocity.
By further assuming that spiral galaxies have a universal mass-to-light ($M/L$) ratio and constant mean surface brightness $(\bar{I})$ we can instead write $L \propto \bar{I}R^2$, from which it follows that
\begin{equation}\label{eq:TF}
    L \propto v_{\mathrm{rot}}^{4}\,.
\end{equation}
It should be noted, though, that the majority of TF surveys find a power law that deviates from the theoretically derived $v^4$. Please consult \cite{Said2023} for a more complete history and overview of the Tully-Fisher relation.

\subsubsection{Fundamental Plane Relation}\label{ssec:FP}
\begin{figure}
    \centering
    \includegraphics[width=0.75\linewidth]{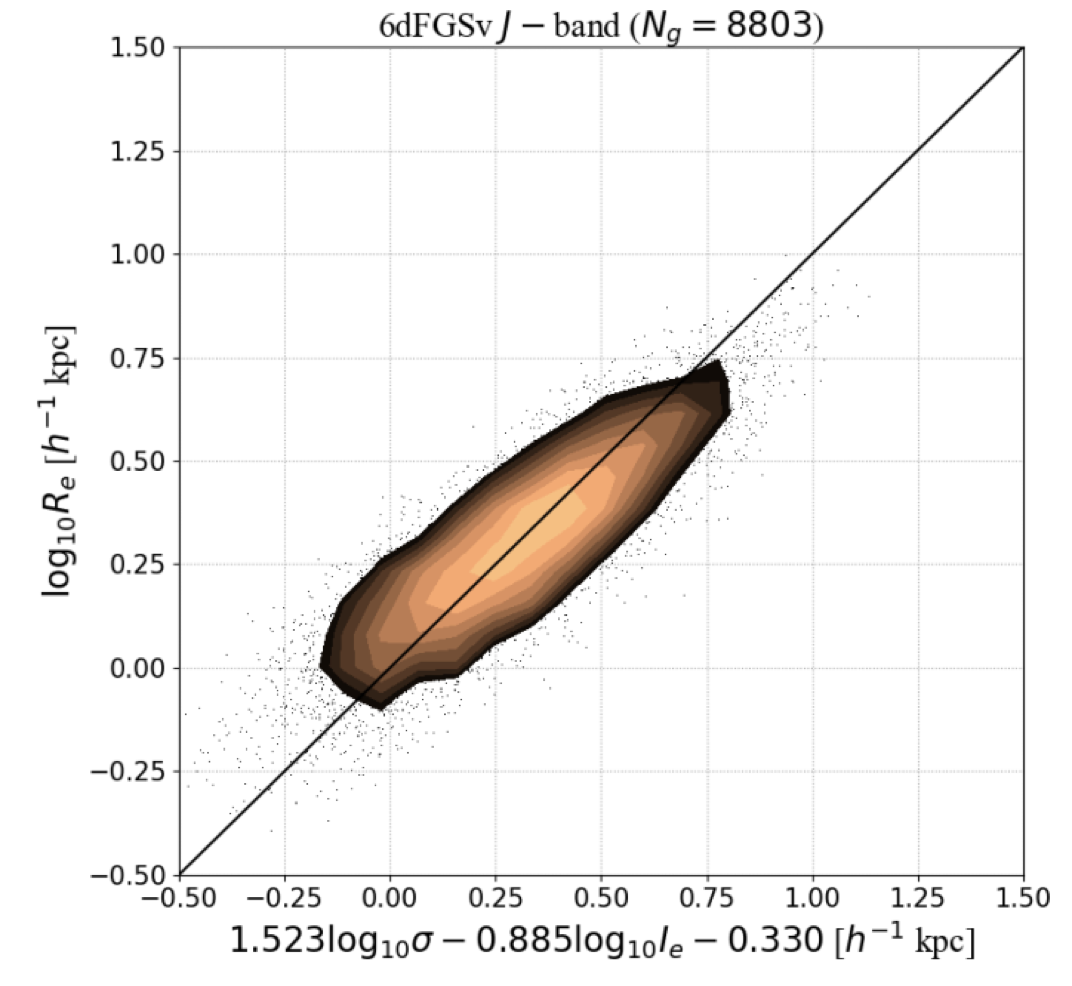}
    \caption{An example of the Fundamental Plane relation, following Eq. \ref{eq:fp}, measured for galaxies in the 6dFGS catalogue. Credit: Said et al. (2020).}
    \label{fig:FPR}
\end{figure}

The Fundamental Plane (FP) relation is a secondary distance indicator applicable to elliptical galaxies, and is the analogue to the TF relation. Elliptical galaxies are not rotation-supported and so the relevant measure of velocity is the velocity dispersion of the central region of the galaxy. The FP relation relates the velocity dispersion to the effective radius of the galaxy, and is based on the Faber-Jackson relation \citep{Faber1967}:
\begin{equation}\label{eq:FJ}
    L \propto \sigma_e^4\,.
\end{equation}
The physical interpretation of the Faber-Jackson relation is also based on the virial theorem and gravitational equilibrium, hence the similarity of Eq. \ref{eq:TF} to \ref{eq:FJ}. The scatter in this relation is roughly twice that of the TF relation. It was shown separately by both \cite{Dressler1987} and \cite{Djorgovski1987} that it is possible to tighten this relation to the point of being comparable to the TF relation if we include surface brightness as a third parameter. Elliptical galaxies can be placed onto a planar region of a three-dimensional parameter space nominally called the Fundamental Plane, described by the effective radius of the galaxy $R_e$ (the radius enclosing half of the total luminosity of the galaxy), the surface brightness interior to the effective radius $I_e$, and the central velocity dispersion $\sigma_v$:
\begin{equation}\label{eq:fp}
    \log_{10}(R_e) = a\,\log_{10}(\sigma_v) + b\,\log_{10}(I_e) + c
\end{equation}
where again the coefficients $a$, $b$, and $c$ describe the slope and zeropoint of the relation and must be calibrated from a sample of galaxies with known distances, see Fig. \ref{fig:FPR} for an example from the 6-degree Field Galaxy Survey (6dFGS) dataset \citep{Said2020}. Similarly to the Tully-Fisher relation, the efficacy of the Fundamental Plane is limited by its intrinsic scatter of around 0.1 dex, which also corresponds to approximately $20\%$ error on distance \citep{Lyndenbell1988}.
\subsection{Type Ia Supernovae}\label{ssec:SNe}
White dwarfs are stellar remnants of stars not massive enough to become a neutron star or a black hole. Typically composed of carbon and oxygen, white dwarfs no longer undergo fusion in their cores and so are instead supported by electron-degeneracy pressure against the force of gravitational collapse. This also means that white dwarfs are some of the densest objects in the universe. The average mass of a white dwarf is $\sim0.6 M_{\odot}$, and the radius of white dwarfs is of order the same as the radius of Earth. There is a mass limit for white dwarfs that, beyond which, they can no longer be supported by electron-degeneracy pressure. This limit, called the Chandrasekhar limit, is $1.44 M_{\odot}$. A white dwarf in a binary system accreting matter from its companion slowly approaches this critical limit, and the temperature in its core will reach a point where nuclear reactions begin again. This sparks a thermonuclear chain reaction, causing the white dwarf to explode in a supernova.

This form of supernova is specifically classified as a Type Ia supernova (SNe Ia). All SNe Ia are caused by similar circumstances and result in the same form of nuclear reaction, thus they are all similarly bright and their luminosities evolve in a characteristic way, evidenced by their light curves. The intrisic luminosity of SNe Ia can be used to infer the distance to the event, giving them the name `standard candles'. This is somewhat of a misnomer, however, as astronomers must apply corrections to each SNe to account for differences in the colour (difference in magnitude of the SNe as measured in two different filters or bands) and stretch (shape of the light curve) between individual events. SNe Ia light curves are fit to a template model to determine the stretch, colour, light curve fit parameters, and amplitude of SN flux, and these parameters are used to estimate the distance modulus
\begin{equation}
    \mu = m - M + \alpha x_1 - \beta c + \gamma G_{\mathrm{host}} - \Delta\mu_{\mathrm{bias}}\,,
\end{equation}
where $m = -2.5\log_{10}(x_0)^2$ is the apparent magnitude, $x_0$ is the SN flux, $x_1$ is the stretch, and $c$ is the colour, while $G$ and $\Delta\mu$ are additional corrections to the distance modulus $\mu = m - M$. $\alpha$, $\beta$, and $\gamma$ are constants determined from the likelihood analysis of all SNe on the Hubble diagram \citep[see][]{DES2024}.

Type Ia SNe were pivotal in determining that the universe is expanding at an accelerated rate, for which Saul Perlmutter, Adam Riess and Brian Schmidt received the Nobel Prize in Physics in 2011, and in addition can also be used to obtain peculiar velocity measurements for their host galaxies. The peak apparent magnitude of a SNe Ia can be used to infer its luminosity distance, and thus the `true' cosmological redshift to the galaxy it resides in. 

As mentioned, though, SNe Ia are secondary distance indicators, and the redshift-distance relationship constructed from supernovae observations must be calibrated with the `local distance ladder', a process by which we use accurately known distances to nearby celestial objects to anchor measurements of objects which are further away. We can determine the distances to the closest stars to Earth using parallax, and use these distances to then determine the distance to primary distance indicators such as Cepheid variable stars. Cepheids are a particular type of variable star whose luminosity varies with time, and the peak luminosity is related to the frequency of the brightness variations. With enough Cepheids, we are able to construct a luminosity-distance relationship to inform us about the distance to more distant Cepheids. If we know that Cepheids stars exist in the host galaxy of a supernova, it follows that we then that we must know the distance to the supernova.

Distances obtained from SNe Ia are considerably more precise than those from the TF or FP relations, with percentage errors typically on the order of $5-10\%$. However, the number of reliably measured SNe distances is to date much smaller than those from either of the galaxy scaling relations. 

In recent years immense effort has gone into producing several supernovae datasets and correcting the data therein for the multitude of systematic uncertainties inherent to measuring supernovae. The most recent datasets are:
\begin{itemize}
    \item Pantheon+ is a compilation of 1701 light curves measured for 1550 unique spectroscopically-confirmed Type Ia supernovae, compiled from 18 different samples and covering redshifts from $z = 0.001$ to $2.26$, used by the SH0ES (Supernovae and H$_0$ for the Equation of State) team in their distance-ladder analysis. Pantheon+ was released in 2022 and is an updated version of the original Pantheon catalogue of supernovae measurements, which contained 1048 unique supernovae. The entire sample is recalibrated and their light curves refitted with a SALT2 model. The legacy of the Pantheon+ sample is in the number of supernovae catalogued at low redshift, ensuring its usefulness as a means of constraining the Hubble constant.
    \item Union3 presents a similar case, consisting of 2087 Type Ia supernovae compiled from 24 datasets. Union3 and Pantheon+ have several datasets in common (sharing $\sim1360$ supernovae) but Union3 follows a different path with regard to their calibration pipeline, treating their errors with Bayesian Hierarchical Modelling, applying different selections to their samples, and fitting their light curves using the SALT3 model, rather than SALT2. 
    \item The Dark Energy Survey (DES) published their Year 5 data release and cosmological analysis in 2024, presenting 1635 Type Ia supernovae in the redshift range $0.10 < z < 1.13$, classified based on their photometry, and also includes an external sample of 194 low redshift supernovae in the range $0.025 < z < 0.10$, for a total of 1829 SNe. The DES sample focuses on the higher redshift range, and far exceeds the number of supernovae beyond $z > 0.5$ presented by either Union3 or Pantheon+.
\end{itemize}
\subsection{Other methods}
There are many other methods that can be used to estimate distances in the local universe that have not been covered here, like gravitational waves, the Tip of the Red Giant Branch, surface brightness fluctuations, the J-region Asymptotic Giant Branch, carbon stars, megamasers, and the globular cluster luminosity function. The reason being that these methods are generally more applicable to measurements of the Hubble constant, and have not seen recent or widespread use in the measurement of peculiar velocities. For more information on any of these methods, please consult the references in Section 8.6.

\section{Peculiar Velocity Surveys}\label{sec:surveys}
The intrinsic scatter inherent to the TF and FP relations corresponds to a roughly $20\%$ error in the distance measurement for a single galaxy. This makes them rather poor tools for determining distances for individual objects. The real power of the galaxy scaling relations is in the large number of galaxies they can be applied to, resulting in a small statistical error across the whole sample. This is one of the primary motivators for surveys dedicated to producing distances using the TF and FP relations, collecting many distances (and therefore peculiar velocities) over a large volume of the sky. Uncertainties in the distance measurements obtained from the TF and FP relations scale with the distance from the observer. A constant scatter in magnitude corresponds to a distance error proportional to the distance, or a constant error in log-distance. Consequently, surveys employing these techniques are restricted to the local universe ($z < 0.15$) where these errors are manageable, and we can get sufficient signal-to-noise in our observations to obtain the velocities needed for either galaxy scaling relation. 

The TF and FP relations have been used to collect distances since the 1980s by various groups, but these datasets were independently gathered and subject to different corrections and methodologies. This meant that any galaxies shared between different catalogues would have systematically different distance estimates, each group essentially deriving their own specific scaling relations. The efforts of the time culminated in the Mark III catalogue of Galaxy Peculiar Velocities, a collection of previous TF datasets of spirals and existing FP samples of ellipticals re-treated so that corrections are applied uniformly. The TF relations derived for each dataset are recomputed accounting for selection biases, and the coefficients are adjusted so that the distances to galaxies common to more than one of the datasets agree. The Mark III catalogue contains approximately 3200 galaxies and has a mean redshift of $z \sim 0.015$. 

In the last twenty years or so the number of wide-field galaxy redshift surveys, and the total number of galaxies they have observed, has dramatically risen. With advances in spectroscopic capabilities and the construction of new instruments to be fitted to existing telescopes, surveys that can cover hemispheres of the sky and measure galaxies in their millions are only becoming more commonplace. Several of these surveys have dedicated programs or subsurveys for the benefit of peculiar velocity cosmology, applying the Tully-Fisher and/or Fundamental Plane relations to thousands or tens of thousands of galaxies.

In chronological order, the most relevant surveys for this review are:
\begin{itemize}
    \item The Spiral Field I-band ++ \citep[SFI++;][]{Springob2007} survey was, at its time of release in 2007, was the largest Tully-Fisher survey published. The SFI++ catalogue contains around 5000 spiral galaxies that have been measured with the TF relation in the I-band, and is largely composed of previous data obtained by the group, including the SFI survey \citep{Giovanelli1994, Giovanelli1995} and the two Spiral Cluster I-band surveys (SCI; \citealt{Giovanelli1997}, and SC2; \citealt{Dale1999}). An erratum was issued in 2009, after it was discovered that a morphological-type correction had not been applied correctly to the peculiar velocity measurements, and this dataset has a median redshift of $\sim 0.02$ and covers most of the sky above the Galactic plane \citep{Springob2009}.
    \item The 6-degree Field Galaxy Survey \citep[6dFGS;][]{Jones2004, Jones2005, Jones2009} generally refers to two subsurveys, the 6dFGS redshift sample (6dFGSz) and the velocity sample (6dFGSv). While the final 6dFGSz data release happened in 2009, the 6dFGSv sample was only made public in 2014 \citep{Springob2014}. 6dFGSv includes almost 9000 early-type galaxies for which the group had Fundamental Plane measurements, after making stringent catalogue cuts based on signal-to-noise and velocity dispersion. Additional cuts were made based on a maximum heliocentric redshift of $z = 0.55$ and on J-band magnitude. 6dFGS covers the entire southern sky at Galactic latitudes $|b| > 10^{\circ}$, also avoiding the Galactic plane.
    \item The Two Micron All-Sky Survey \citep[2MASS;][]{Skrutskie2006} Tully-Fisher Survey \citep[2MTF;][]{Hong2019} is a catalogue of Tully-Fisher distances for bright spiral galaxies contained in the 2MASS Redshift Survey \citep[2MRS;][]{Huchra2012}, obtained with photometry from 2MASS as well as HI data from the Arecibo Legacy Fast ALFA \citep[ALFALFA;][]{Giovanelli2005, Haynes2018} survey. The final 2MTF data release was published in 2019 and contained distance measurements for around 2000 galaxies out to a redshift of $\sim 0.03$ and a median redshift of $\sim 0.016$. Previous surveys had avoided the Galactic plane due to extinction effects, this means that large amounts of structure in the local universe are simply unaccounted for. 2MTF uses infrared data from 2MRS to reduce these effects and probes closer to the plane at $|b| > 5^{\circ}$, whilst maintaining a near-uniform coverage of the rest of the sky. 
    \item The Sloan Digital Sky Survey Peculiar Velocity sample \citep[SDSS PV;][]{Howlett2022} is currently the largest single homogeneous catalogue of peculiar velocities that has been published, containing over 34000 FP distances for early-type galaxies in the SDSS dataset. The catalogue, published in 2022, extends our understanding of the peculiar velocity field to redshift $z = 0.1$, and the sample has an effective redshift of $z \sim 0.07$. The SDSS PV sample complements the 6dFGSv sample very well. SDSS PV covers the northern sky while 6dFGSv covers the south, the samples contain similar numbers of galaxies within the redshift limit of 6dFGS, although 6dFGS covers twice the area, and SDSS PV extends into the higher redshift regime.
    \item The Cosmicflows compilation has gone through several iterations since the first version of the catalogue was released in 2008. The most recent one, Cosmicflows-4 \citep[CF4;][]{Tully2022}, was made public in 2022, and comprises nearly 56000 galaxies measured using eight methodologies. The largest contributions to this total come from TF and FP samples, while type Ia supernovae fill out most of the rest. The Tully-Fisher sample contains contributions from SFI++ and 2MTF among other sources of TF distances, while the Fundamental Plane sample contains data mainly from 6dFGSv and SDSS PV. Other methods that complete the sample include core-collapse supernovae and surface brightness fluctuations. CF4 expanded on Cosmicflows-3 \citep{Tully2016} by increasing the total amount of TF distances and focusing on expanding the sky coverage of the dataset into the northern sky. While the rest of the surveys discussed here are homogeneous, CF4 is notable for being a heterogeneous compilation of data from different groups using separate methodologies. CF4 has coverage in both the northern and southern sky, whilst not being as uniform as previous surveys due to its nature as a compilation of other datasets.
\end{itemize}

\section{Using Peculiar Velocities for cosmology}\label{sec:cosmo}
We have covered the mathematics of the peculiar velocity field, and how peculiar velocities measured at a point in this field are induced by the gravitational influence of matter overdensities in the surrounding environment. Further, we have explored the various methods of probing the velocity field at specific points by obtaining redshifts and redshift-independent measures of distance for tracers of the velocity field, and how major collaborations are producing large samples of such measurements. Now, we will look into the methods used to produce cosmological constraints using these peculiar velocity datasets. For additional information on the velocity and density fields, please see the review by \cite{Strauss1995} and references therein.
\subsection{Bulk flows}
Earlier we described how the measurement of many peculiar velocities enables us to chart out the velocity flows of galaxies in the local universe as they are pulled in different directions under gravity, in a process known as cosmography. If we were to take the average of these motions we would arrive at a measurement of the `bulk flow', an estimate of the amplitude and direction of the movement of all of the matter we've averaged over. Averaging over different volumes, we can measure the bulk flow of the matter in those volumes, and build up a picture of how matter in different regions of the local universe is, on average, moving. Mathematically, and in the simplest terms, the bulk flow $\textbf{B}$ in a given volume $V$ is given by
\begin{equation}
    \textbf{B} = \frac{1}{V}\int_V \textbf{v}(r)\,d^3r
\end{equation}
where we have integrated over all of the three-dimensional peculiar velocities $v(r)$ within the volume. Significant effort has gone into measuring the bulk flow of the `Local Group' (LG), the group of galaxies containing the Milky Way among others. The LG bulk flow gives us a measure of the amplitude and direction of the velocity of our local universe that can be compared to the $\Lambda$CDM prediction, which is a distribution centred on zero with some width. If the measured bulk flow lies outside of the predicted distribution then this may present a potential avenue for disproving, or suggesting additional physics beyond, $\Lambda$CDM.

\subsection{Two-point statistics}
To this point we have understood the matter density field to have evolved under the linear gravitational instability model of structure formation, where the initial fluctuations in the density field originated as quantum fluctuations that grew to astronomical proportions during inflation. One consequence of this model is that we have assumed that the initial fluctuations constitute a Gaussian random field, and therefore the statistics of the field can be completely characterised by two-point statistics. Two more assumptions that we must make are related to the cosmological principle, isotropy and homogeneity. Under isotropy, there are no special directions in the universe, and so any statistics we measure should be unchanged when the considered positions in the field are rotated. Under homogeneity, we are positing that there are no special places in the universe, and so any statistics we measure should be identical under translation. This means that if we average over a large enough volume, we should recover the true underlying properties of the field. From these assumptions we finally arrive at the two-point correlation function and the power spectrum.

The correlation function is a statistic that describes the structure of fluctuations in the distribution as a function of scale. Following \cite{Peebles1980}, and borrowing some stylistic choices from \cite{Huterer-book}, let us first consider the probability of finding a point-like object is some infinitesimal volume $\delta V$, where the mean number density is $n$
\begin{equation}
    \delta P = n\delta V\,.
\end{equation}
Now consider the joint probability of finding an additional point-like source, one in the infinitesimal volume element $\delta V_1$ and the other in $\delta V_2$
\begin{equation}\label{eq:2p_prob}
    \delta P = n^2 \delta V_1 \delta V_2 \left[1 + \xi(r_{12})\right]
\end{equation}
where $\xi$ is the two-point correlation function, and determines the excess probability of finding the second point a distance of $r_{12}$ away from the first, as compared to a uniform random Poisson point process where both probabilities are independent and $\delta P = n^2\delta V_1\delta V_2$. In the case of a uniform Poisson process $\xi \equiv 0$, if the two positions are correlated then $\xi > 0$, if the two positions are anti-correlated then $-1 \leq \xi < 0$.
The value of $\xi$ cannot be any lower than -1, else Eq. \ref{eq:2p_prob} would imply negative probability. Assuming that we observe the first object with probability $n\delta V_1$, the conditional probability of finding the second particle in $\delta V_2$ is
\begin{equation}
    \delta P(2|1) = n\delta V_2\left[1 + \xi(r_{12})\right]\,.
\end{equation}
Generalising to a randomly chosen object in our ensemble, the probability of finding a second object a distance of $r$ away in the volume element $\delta V$ is
\begin{equation}
    \delta P = n\delta V\left[1 + \xi(r)\right]\,,
\end{equation}
and so expected number of neighbours of a random object in the volume, within a distance $\textbf{r}$ from that object is
\begin{equation}
    \langle N \rangle = nV + n\int_0^r \xi(r) dV\,.
\end{equation}
The two-point correlation function of a continuous density field $\rho(\textbf{x})$ can then be written as
\begin{equation}
    \xi(r) = \frac{\langle \left[ \rho(\textbf{x} + \textbf{r}) - \langle \rho \rangle\right] \left[\rho(\textbf{x}) - \langle \rho \rangle\right]\rangle}{\langle \rho \rangle^2} = \langle \delta(\textbf{x})\,\delta(\textbf{x} + \textbf{r}) \rangle\,.
\end{equation}
We have multiplied an overdensity at position $\textbf{x}$ with another ovedensity at position ($\textbf{x} + \textbf{r}$), recall Eq. \ref{eq:contrast} where we are now ignoring any time dependencies. We also average over all positions \textbf{x}, under the assumption of homogeneity, and all directions of \textbf{r}, under the assumption of isotropy. 

The power spectrum $P(k)$ is the Fourier counterpart to the two-point correlation function $\xi(r)$, and represents the distribution of fluctuations in the density field in Fourier space as a function of the wavenumber $k$,
\begin{equation}
    \xi(r) = \frac{1}{2\pi^2}\int P(k)\,k^2 j_0(kr)\,dk
\end{equation}
or
\begin{equation}
    P(k) = \int \xi(r)\,e^{-i\,\textbf{kr}}d^3\textbf{r}\,,
\end{equation}
where $j_0(kr) = sin(kr)/(kr)$ is the zeroth-order spherical Bessel function.
We can also define the power spectrum $P(k)$ as the ensemble average of the squared amplitude of all of the Fourier modes contained in the field
\begin{equation}
    P(k) = \langle | \delta(\textbf{k}) |^2 \rangle\,.
\end{equation}
We have to this point described the two-point correlation function in its simplest form, correlating two positions in the density field, but we can also define it explicitly in terms of the galaxy power spectrum, as a function of separation:
\begin{equation}
    \xi_{gg}(\textbf{r}) = \int\frac{dk}{2\pi^2}P_{gg}(k) \,k^2 j_0(kr)
\end{equation}
where the galaxy power spectrum, $P_{gg}(k) = b^2P(k)$, does not depend on the direction of \textbf{k} when neglecting the effects of redshift-space distortions, because of isotropy.

\begin{figure}
    \centering
    \includegraphics[width=0.5\linewidth]{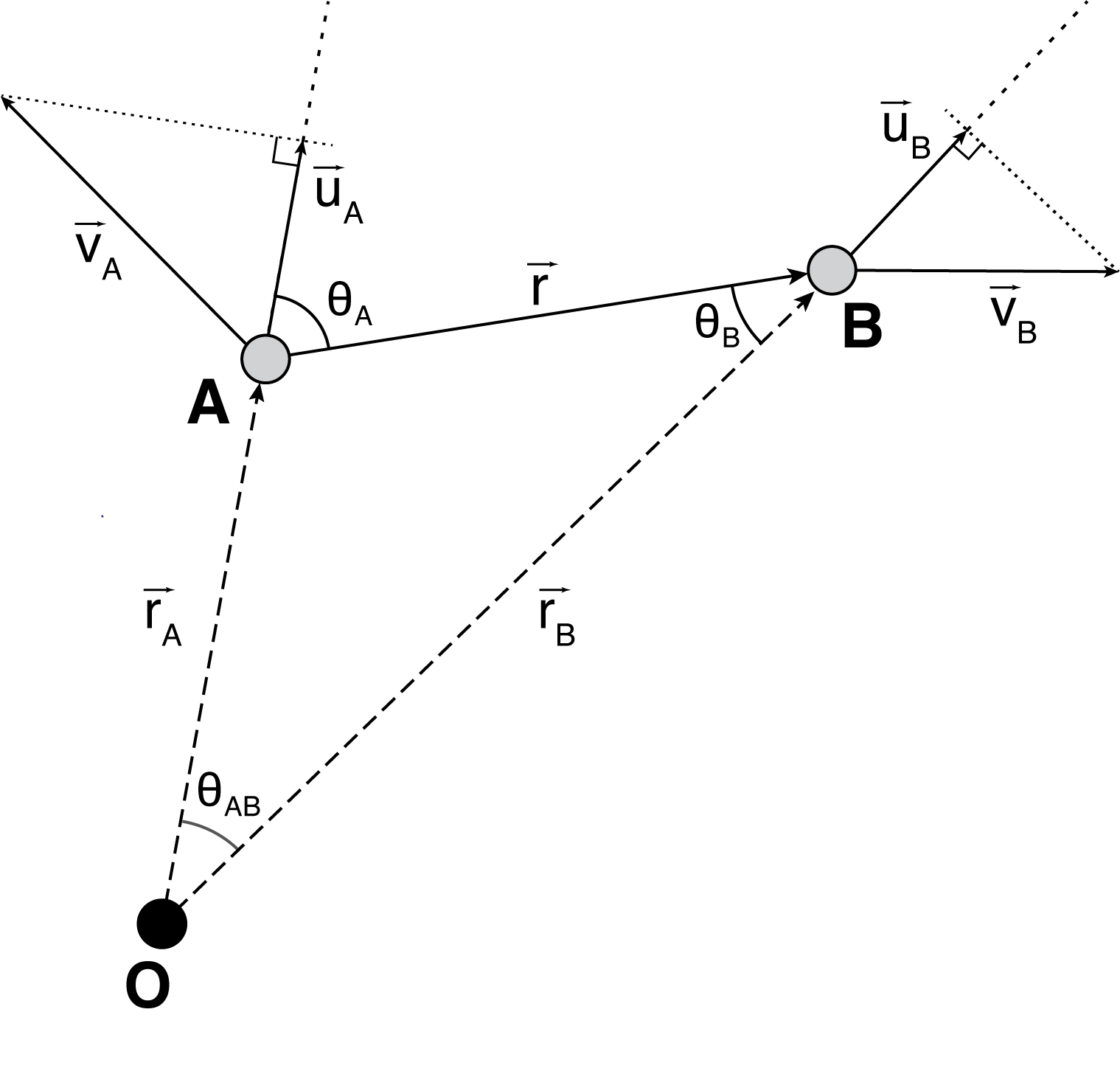}
    \caption{A simple view of the geometry describing the system of a pair of galaxies $A$ and $B$, separated by some vector $\vec{r}$, in relation to an observer $O$. The three-dimensional velocities are given by $\vec{v}_i$ and the line-of-sight component by $\vec{u}_i$. The vectors describing the separation between the two objects and the observer are given by $\vec{r}_i$ and the angles between the three vectors by $\theta_i$.}
    \label{fig:pair-geom}
\end{figure}

We can also describe the velocity power spectrum, however because the velocity is a vector quantity we must use a different approach. Instead, we write the two-point statistics of the peculiar velocity field as a three-dimensional correlation tensor, with the general form:
\begin{equation}
    \Psi_{ij}(r_A, r_B) = \langle v_i(r_A)\,v_j(r_B) \rangle
\end{equation}
where $r_A$ and $r_B$ are two spatial positions, $i$ and $j$ represent components of the 3D peculiar velocity, and $\langle ... \rangle$ represents the ensemble average across many realisations, for more information see \cite{Gorski1988}. Assuming that the velocity field is irrotational (curl-free), homogeneous, isotropic, and perturbations in the velocity field are linear, we can rewrite the tensor in the form:
\begin{equation}\label{eq:3dvelcorr}
    \Psi_{ij}(r) = \left[\Psi_{\parallel}(r) - \Psi_{\perp}(r)\right]\left(\frac{r_ir_j}{r^2}\right) + \Psi_{\perp}(r)\,\delta_{ij}^K
\end{equation}
where $\delta_{ij}^K$ is the Kronecker delta function, and $\Psi_{\parallel}(r)$ and $\Psi_{\perp}(r)$ are two isotropic functions given by:
\begin{equation}
    \Psi_{\parallel}(r) = \int\frac{dk}{2\pi^2}\,k^2\,P_{vv}(k)\left[j_0(kr) - \frac{2j_1(kr)}{kr}\right]\,,
\end{equation}
and,
\begin{equation}
    \Psi_{\perp}(r) = \int\frac{dk}{2\pi^2}\,k^2P_{vv}(k)\frac{j_1(kr)}{kr}\,,
\end{equation}
where $P_{vv}(k) = \frac{a^2H^2f^2}{k^2}\,P_{m}(k)$ is the velocity power spectrum,  and $j_1(x) = \sin{(x)}/x^2 - \cos{(x)}/x$ is the first-order spherical Bessel function. $\Psi_{\parallel}$ and $\Psi_{\perp}$ are the radial and transverse correlation functions of the three-dimensional peculiar velocity field respectively. Up to this point we have been working with the three-dimensional velocity, which we cannot directly measure, and so we must recast the velocity correlation function in terms of the observable line-of-sight velocities \textbf{$u$}, turning Eq. \ref{eq:3dvelcorr} into
\begin{equation}\label{eq:los_velcor}
    \langle u_A(\textbf{x})\,u_B(\textbf{x}+\textbf{r}\rangle = \Psi_{\perp}(r)\cos\theta_{AB} + \left[\Psi_{\parallel}(r) - \Psi_{\perp}(r)\right]\cos\theta_A\cos\theta_B
\end{equation}
where $\cos\theta_{AB} = \hat{\textbf{r}}_A.\hat{\textbf{r}}_B$, $\cos\theta_A = \hat{\textbf{r}}.\hat{\textbf{r}}_A$, and $\cos\theta_B = \hat{\textbf{r}}.\hat{\textbf{r}}_B$. For reference, we show a simple view of the geometry of a given pair of galaxies A and B in relation to an observer O in Fig. \ref{fig:pair-geom}.
Enforcing a local flat-sky approximation ($\theta_{AB} = 0, \theta_A = \theta_B = \theta$), Eq. \ref{eq:los_velcor} simplifies to
\begin{equation}\label{eq:3dvelcor}
    \langle u_A\,u_B\rangle = \Psi_{\perp}(r) + \left[\Psi_{\parallel}(r) - \Psi_{\perp}(r)\right]\cos^2\theta\,.
\end{equation}
Physically, the two-point correlation function is composed of a linear combination of these two isotropic functions, and we measure the total correlation differently depending on the angle of separation between the two sources, relative to us. If the pair is separated by a vector parallel to our line-of-sight, then $\theta = 0^{\circ} $ and $\langle u_Au_B \rangle = \Psi_{\parallel}(r)$. If the pair is separated by a vector perpendicular to our line-of-sight, then $\theta = 90^{\circ} $ and $\langle u_Au_B \rangle = \Psi_{\perp}(r)$, stressing that this is in the flat-sky approximation ($\theta_{AB} = 0$) as opposed to the scenario depicted in Fig. \ref{fig:pair-geom}. It does bear mentioning that the velocity two-point correlations are technically momentum (density-weighted velocity) correlations \citep{Park2000, Howlett2019}. To measure a velocity at a point in the velocity field we require some tracer of the field, namely a galaxy. In the absence of any galaxies we have no tracers and so suffer from `zero-valued' regions, where the velocity field must arbitrarily be set to zero. The momentum field, given by $\rho(r) = (1 + \delta_g(r)) \cdot v(r)$, does not suffer from these arbitrary zero-valued regions.

It was \cite{Gorski1989} that introduced two estimators of the line-of-sight velocity auto-correlation function that depend only on the radial component of the galaxy peculiar velocity. These statistics, named $\psi_1$ and $\psi_2$, are defined as:
\begin{equation}\label{eq:psi1}
    \psi_1(r) = \frac{\sum u_A u_B \cos\theta_{AB}}{\sum \cos^2\theta_{AB}}\,,
\end{equation}
\begin{equation}\label{eq:psi2}
    \psi_2(r) = \frac{\sum u_A u_B \cos\theta_A \cos\theta_B}{\sum \cos\theta_{AB} \cos\theta_A \cos\theta_B}\,,
\end{equation}
where the sums are over all galaxy pairs separated by some fixed distance $r$. \cite{Gorski1989} also showed that the models for these estimators can be written as linear combinations of the two isotropic functions of the three-dimensional velocity field:
\begin{equation}
    \langle \psi_1(r) \rangle = \mathcal{A}(r)\Psi_{\parallel}(r) + \left(1 - \mathcal{A}(r)\right)\Psi_{\perp}(r)\,,
\end{equation}
\begin{equation}
    \langle \psi_2(r) \rangle = \mathcal{B}(r)\Psi_{\parallel}(r) + \left(1 - \mathcal{B}(r)\right)\Psi_{\perp}(r)\,,
\end{equation}
where,
\begin{equation}
    \mathcal{A}(r) = \frac{\sum \cos\theta_{AB} \cos\theta_A \cos\theta_B}{\sum \cos^2\theta_{AB}}\,,
\end{equation}
\begin{equation}
    \mathcal{B}(r) = \frac{\sum \cos^2\theta_A \cos^2\theta_B}{\cos\theta_{AB} \cos\theta_A \cos\theta_B}\,.
\end{equation}

$\mathcal{A}$ and $\mathcal{B}$ contain information about the geometry of the survey and measure the contributions of $\Psi_{\perp}$ and $\Psi_{\parallel}$ to $\psi_1$ and $\psi_2$, respectively. It has been shown that $\psi_1$ is somewhat more robust to sample fluctuations than $\psi_2$, as Eq. \ref{eq:psi1} is always positive and non-vanishing while Eq. \ref{eq:psi2} can be either positive or negative.

Alternatively, another correlation that we can measure with velocity data is the mean pairwise velocity, the relative velocity of pairs of galaxies at some separation $r$ measured along the vector of separation. In its simplest form, the mean pairwise velocity is given by:
\begin{equation}
    v_{12}(r) = \langle (\textbf{u}_A - \textbf{u}_B)\cdot\hat{\textbf{r}} \rangle = \frac{\langle (\textbf{u}_A - \textbf{u}_B)(1 + \delta_A)(1 + \delta_B)\rangle}{1 + \xi(r)}\,
\end{equation}
where $\langle...\rangle$ indicates that we are averaging over all pairs, and we have related $v_{12}(r)$ to the two-point correlation function $\xi(r)$ via the pair conservation equation \citep{Peebles1980}. $v_{12}(r)$ is an equivalent measurement to the galaxy-velocity cross-correlation function, which correlates the velocity of one galaxy along the separation vector towards another (treating the second galaxy purely as a tracer of density). This cross-correlation between the overdensity and galaxy velocity fields is given by
\begin{equation}
    \langle \delta(\textbf{x})\,\textbf{v}(\textbf{x} + \textbf{r}) \rangle = \hat{r}\,\xi_{gv}(r)
\end{equation}
where,
\begin{equation}
    \xi_{gv}(r) = -\frac{Hafb}{2\pi^2}\int dk\,k\,P_m(k)\,j_1(kr)\,,
\end{equation}
\citep[see][]{Fisher1995, Adams2017}.

Rewriting this as a cross-correlation between overdensity and line-of-sight velocity gives:
\begin{equation}
    \langle \delta(\textbf{x})\,\textbf{u}(\textbf{x} + \textbf{r}) \rangle = \xi_{gv}(r) \cos\theta_B\,,
\end{equation}
and thus an estimator of the cross-correlation can be defined as
\begin{equation}
    \psi_3(r) = \frac{\sum u_B \cos\theta_B}{\cos^2\theta_B}\,,
\end{equation}
with a model given by:
\begin{equation}
    \langle \psi_3(r) \rangle = \frac{\sum \langle \delta u_B \rangle \cos\theta_B}{\cos^2\theta_B} = \xi_{gv}(r)\,,
\end{equation}
\citep[see][]{Turner2021}.

\subsection{Velocity field reconstructions}
There is linked cosmological information in the density and velocity fields. Various approaches exist to simultaneously analyse this information within the context of linear theory. We can either convert the density field to a model velocity field, or else analyse the joint two-point statistics of all of the fields. Many approaches have been proposed in the literature to do this, and in linear theory these should give equivalent results. 
Perhaps the earliest methods of constraining cosmology with peculiar velocities involved direct comparisons between the galaxy density field and the peculiar velocity field, again see \cite{Strauss1995} for more information. We have established, via Eqs. \ref{eq:pv-grav} and \ref{eq:bias}, that the velocity and matter density fields are intimately linked via linear perturbation theory, and by substituting the matter density field for the galaxy density field under the assumption of a linear scaling between the two, we can directly link the distribution of galaxies to the velocity field, related by the parameter $\beta = f/b$. There exists two prevailing methods in this linear perturbation theory regime, density-density comparisons and velocity-velocity comparisons.

In density-density comparisons, directly measured peculiar velocities are used to simultaneously derive a three-dimensional velocity field and mass density field. The mass density field is then directly compared to the galaxy density field as measured from galaxy redshift surveys in order to constrain $\beta$. The most prominent example of this is the POTENT method \citep{Bertschinger1989, Dekel1990}, and its use in conjunction with the peculiar velocities in the Mark III velocity catalogue to produce a three-dimensional velocity field and, via gravitational instability theory, reconstruct the underlying mass density field. 

Velocity-velocity comparisons flip this method on its head. By instead estimating the galaxy density field from galaxy redshift surveys and translating that to a reconstruction of the velocity field, assuming some value of $\beta$, we can compare the reconstructed velocities to directly measured velocities. By performing some form of fit to $\beta$, where we are looking for the best agreement between the inferred and observed velocities, we can constrain the parameter. One of the first such approaches was the VELMOD method \citep{Willick1997VELMOD}, a maximum-likelihood method that takes in Tully-Fisher observables and a galaxy redshift, determines the likelihood of observing those values given the measured redshift, and maximises the probability over the dataset given a model of the velocity field.

\begin{figure}
    \centering
    \includegraphics[width=0.75\linewidth]{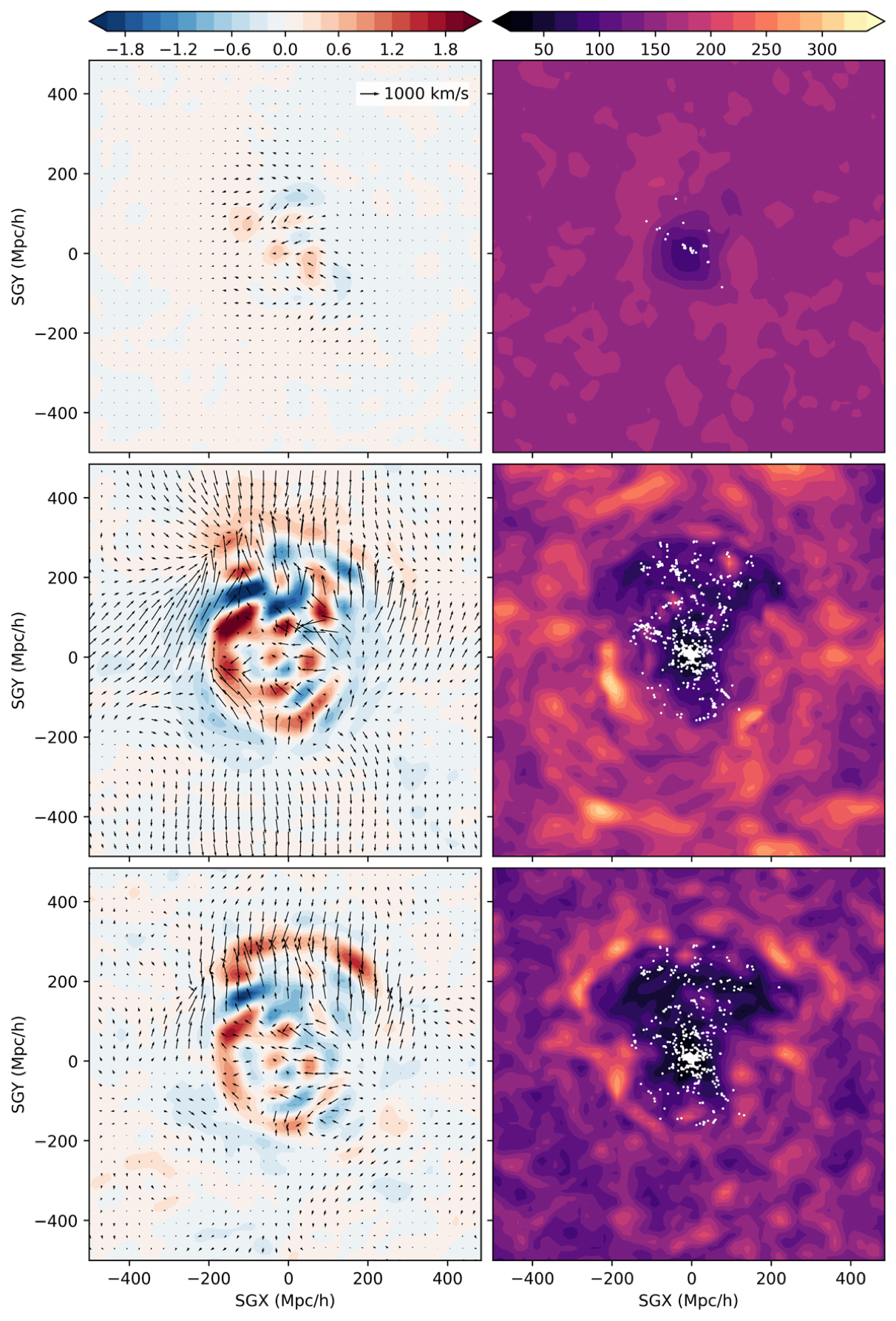}
    \caption{Left column: Three different reconstructions of the three-dimensional overdensity ($\delta$) and velocity fields produced using data from Cosmicflows-4, projected onto supergalactic coordinates in the SGX-SGY plane. Axes are chosen so that the observer is located near the centre of the box. Arrows show the magnitude and direction of the velocity field at that location. Right column: the standard deviation in the velocity field, in units of km s$^{-1}$, for the corresponding panel in the left column. The white dots represent a subsample of galaxies in the CF4 sample. The top row depicts the CF4 supernovae sample, the middle row depicts the ungrouped CF4 galaxies, and the bottom row depicts the grouped CF4 galaxies. Credit: Courtois et al. (2023).}
    \label{fig:cf4-recon}
\end{figure}

Density-density methods suffer from a reliance on peculiar velocity data as the source for their reconstructions, which can be quite noisy and sparse on the sky (the Mark III catalogue, for example, only containing $\sim3200$ galaxies), leading to less reliable results. Velocity-velocity methods on the other hand base their reconstructions on galaxy redshift survey data which is typically measured with much more accuracy and is much more complete across the sky, and it is these methods which prevail today. We show one such example of these reconstructions from the Cosmicflows-4 team in Fig. \ref{fig:cf4-recon}, showing the overdensity field and velocity field reconstructed from supernovae, individual galaxies, and galaxy groups.

\subsection{Maximum-likelihood estimators}

Rather than modelling the velocity field through manipulation of the galaxy density field, or vice versa, we can instead model the covariance between the two fields analytically. This method assumes that both fields are correlated samples drawn from multivariate Gaussian distributions, which have means of zero. This allows us to produce a likelihood function that can be used within a maximum-likelihood (ML) methodology, meaning that the velocity and galaxy density fields can be modelled simultaneously in a self-consistent manner.

The likelihood is defined as usual, being the probability of observing our data given a model
\begin{equation}
    \mathcal{L} = p(\Delta|\phi) = \frac{1}{\sqrt{(2\pi)^N|\mathcal{C}(\phi)|}}\exp\left(-\frac{1}{2}\Delta^T\mathcal{C}(\phi)^{-1}\Delta\right)
\end{equation}
where $\Delta$ is the vector containing the galaxy overdensities and peculiar velocities, $\phi$ is the vector containing model parameters, and $\mathcal{C}$ is the covariance between each element of $\Delta$. The diagonal elements of the covariance matrix represent variance due to the assumed cosmology, which may change the amplitude of clustering in the local universe, while the off-diagonal terms represent the covariance between individual velocity measurements due to them being created from the same density field and sharing Fourier modes.
\section{State of the Field}\label{sec:current}

The field of peculiar velocity cosmology has come on a long way in the last two decades. The size of the datasets we are now producing have increased by orders of magnitude, and we are approaching the point where we will have covered the majority of the sky out to a redshift of at least $z = 0.10$. This is mirrored by the progress made in techniques used to extract cosmological information from these next-generation surveys.
\subsection{Current results}
In the field of two-point statistics, recent results have been reported using data from surveys such as 6dFGS, Cosmicflows, and SDSS. Current correlation function analyses work within linear theory, but capture information from RSD by performing multipole expansions of the two-point auto-correlation functions and the two-field cross-correlation function. For an example of these, see \cite{Turner2023}. The velocity power spectrum has historically been used to constrain the growth rate, but more recent efforts focus on the momentum power spectrum instead, see \cite{Qin2019}.

Velocity field reconstructions continue to be a popular method of utilising peculiar velocity information, and several software packages and new ways of approaching these techniques have been published in the last few years. Cosmicflows-4 published their reconstructions, as well as estimates of the bulk flow, $f\sigma_8$, and $H_0$ in \cite{Courtois2023}. \cite{Boruah2020} compared SNe Ia peculiar velocities to the reconstructed velocity field derived using galaxy redshifts from 2MRS, 6dFGS and SDSS, among others. Velocity correlation analyses have produced measurements of the growth rate with $15-20\%$ measurement error, consistent with $\Lambda$CDM. Some reconstruction methods have published estimates with $5-10\%$ measurement error, some of which claim a disagreement with $\Lambda$CDM with $2-3\sigma$ significance. Peculiar velocity measurements of growth are already competitive with RSD growth measurements in the local universe, and will improve rapidly in the future.

Several estimates of the bulk flow have been made with the Cosmicflows-4 catalogue, utilising different methods, for an example of one of these see \cite{Whitford2023}. What is interesting about current bulk flow measurements is that there is seemingly a split between authors as to whether measurements of the bulk flow are in agreement with $\Lambda$CDM or are in some amount of tension with it. One trend observed by \cite{Whitford2023} is that measurements in tension with $\Lambda$CDM tend to measure the bulk flow on larger scales than the measurements that are in agreement. These disagreements have existed within this field more some time now, and more work is needed to determine if this is another indication of possible missing physics, or is a symptom of undiagnosed systematics or poorly understood model dependencies.

The maximum-likelihood method has seen more uptake in supernova cosmology groups, thanks in part to the software package \textbf{\sc{flip}}. An example of field level inference used in this context is \cite{Carreres2023}. The work of \cite{Adams2020}, applying the maximum-likelihood method to the 6dFGS catalogue, remains one of the more prominent examples of the method.

\begin{figure}
    \centering
    \includegraphics[width=\linewidth]{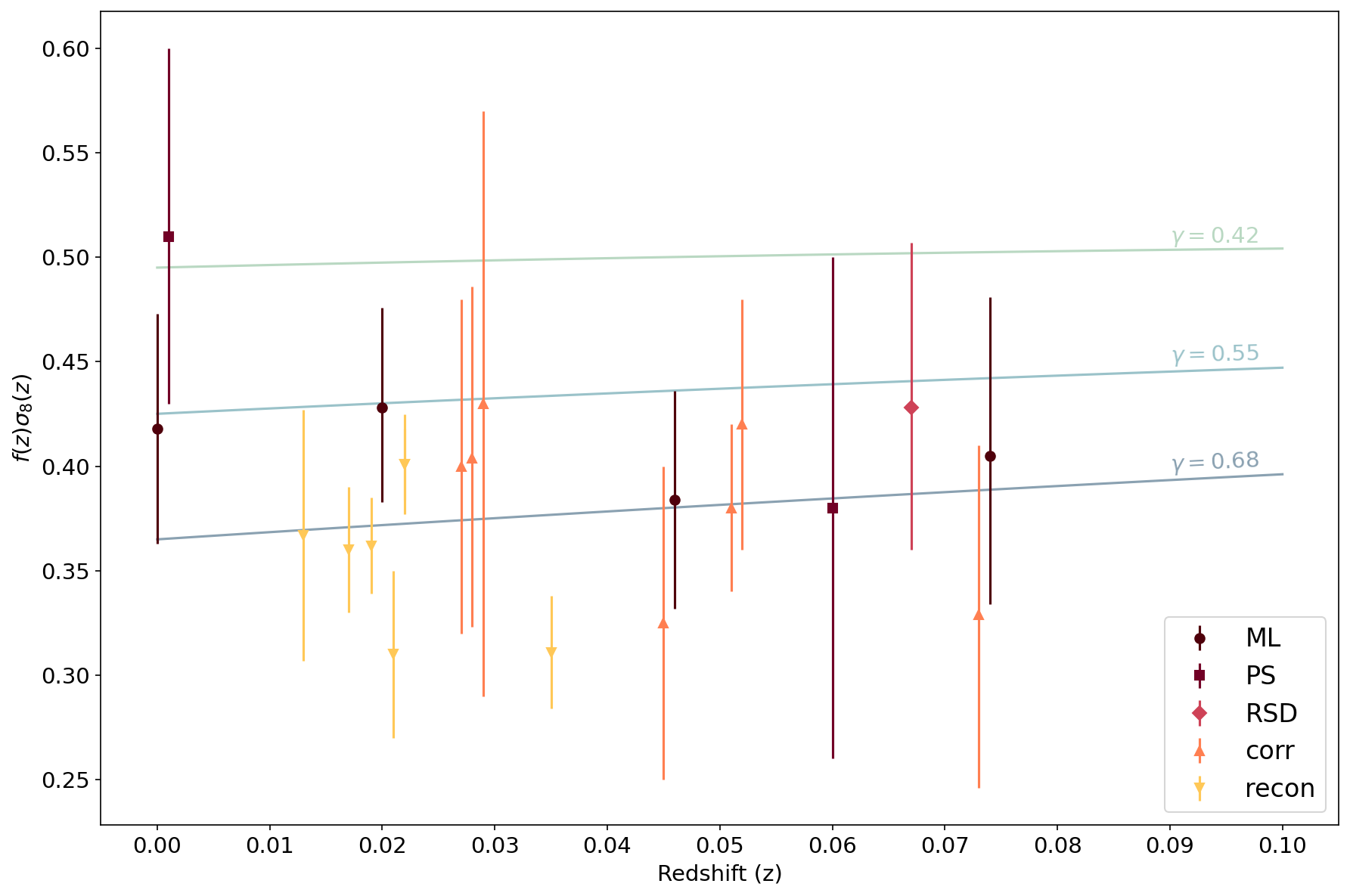}
    \caption{A selection of measurements of the combined parameter $f\sigma_8$ at different redshifts. The three horizontal lines represent three different parameterisations of gravity, indicated by different values of $\gamma = \{$$0.42, 0.55\,$(GR)$,0.68$\}, otherwise assuming a Planck 2018 cosmology. The various measurements are differently-coloured and represented by different shapes depending on the methodology used by the authors. Black circles indicate a maximum-likelihood method (ML), brown squares indicate a power spectrum method (PS), red diamonds indicate a redshift-space distortion method (RSD), orange triangles indicate a method employing two-point correlation functions (corr), and yellow upside-down triangles indicate a method employing some form of velocity-field reconstruction (recon).}
    \label{fig:fs8-comp}
\end{figure}

Fig. \ref{fig:fs8-comp} depicts a collection of recent and prominent measurements of the growth rate using the techniques and methods discussed above, presented as a function of redshift against different $\gamma$-parameterisations of gravity in an otherwise Planck 2018 cosmology. The different methodologies have been separated by colour and shape for clarity. In general, there is a broad agreement between all of the methods and with a $\Lambda$CDM cosmology, but with a slight preference for a higher value of $\gamma$ more attributable to modified gravity scenarios such as DGP gravity. This may suggest a suppression of the growth rate of structure \citep{Nguyen2023}. All results in Fig. \ref{fig:fs8-comp} are also presented in Table \ref{tab:fs8-results}.

\subsection{Future surveys}
The Dark Energy Spectroscopic Instrument (DESI) is installed on the 4m Mayall telescope at Kitt Peak National Park, Arizona and is, at the time of writing, three years into a five year observing strategy targeting 14000 square degrees in the northern hemisphere. It has a 3 degree field of view and a focal plane that holds 5000 optical fibres, enabling the instrument to obtain spectra for 5000 individual galaxies at once. The DESI peculiar velocity (DESI PV) survey is a `spare-fibre' survey, capitalising on the design of DESI to make use of fibres that would otherwise not be required for the main DESI survey. The DESI PV survey will be the first major survey to measure both FP and TF distances using the same instrument, utilising a new way of using optical data to measure the TF relation rather than HI. Based on recent target selections released for the DESI PV survey, they will obtain $\sim 133000$ FP distances and $\sim 53000$ TF distances \citep{Saulder2023}.

The Wide-field ASKAP L-band Legacy All-sky Blind surveY (WALLABY) is a pathfinder survey for the Square Kilometer Array, a radio telescope located across two locations: the low frequency component in Murchison, Western Australia and the mid frequency component at Meerkat National Park, South Africa. WALLABY differs from the other surveys discussed in that it purely observes in the radio, focused on the 21cm line of neutral hydrogen. The WALLABY survey has been observing since 2022, and during its five year observing run will cover 14000 square degrees in the southern hemisphere, over a redshift range $0 < z < 0.10$ with 30-arcsecond resolution. As a radio survey, WALLABY is well-suited to utilising the TF relation to measure distances for tens of thousands of galaxies.

The 4MOST (4-metre Multi-Object Spectroscopic Telescope) Hemisphere Survey (4HS) is a future survey that will measure spectroscopy and redshifts for approximately six million galaxies over 17000 square degrees in the southern hemisphere. One of the major aims of 4HS is to measure galaxies with high ($\sim95\%$) and unbiased completeness out to $z < 0.15$. 4HS also makes use of the structure of 4MOST to achieve this excellent coverage over such a large area, and will be able to probe further into the Zone of Avoidance than its contemporary optical surveys. One of the subsamples to be measured by 4HS is designed specifically to enhance peculiar velocity cosmology. Combining the PV subsample with the main 4HS survey, 4HS will measure velocity dispersions for over 400000 early-type galaxies within this redshift range, which will be the largest sample of FP distances ever produced. 

\begin{figure}
    \centering
    \includegraphics[width=0.75\linewidth]{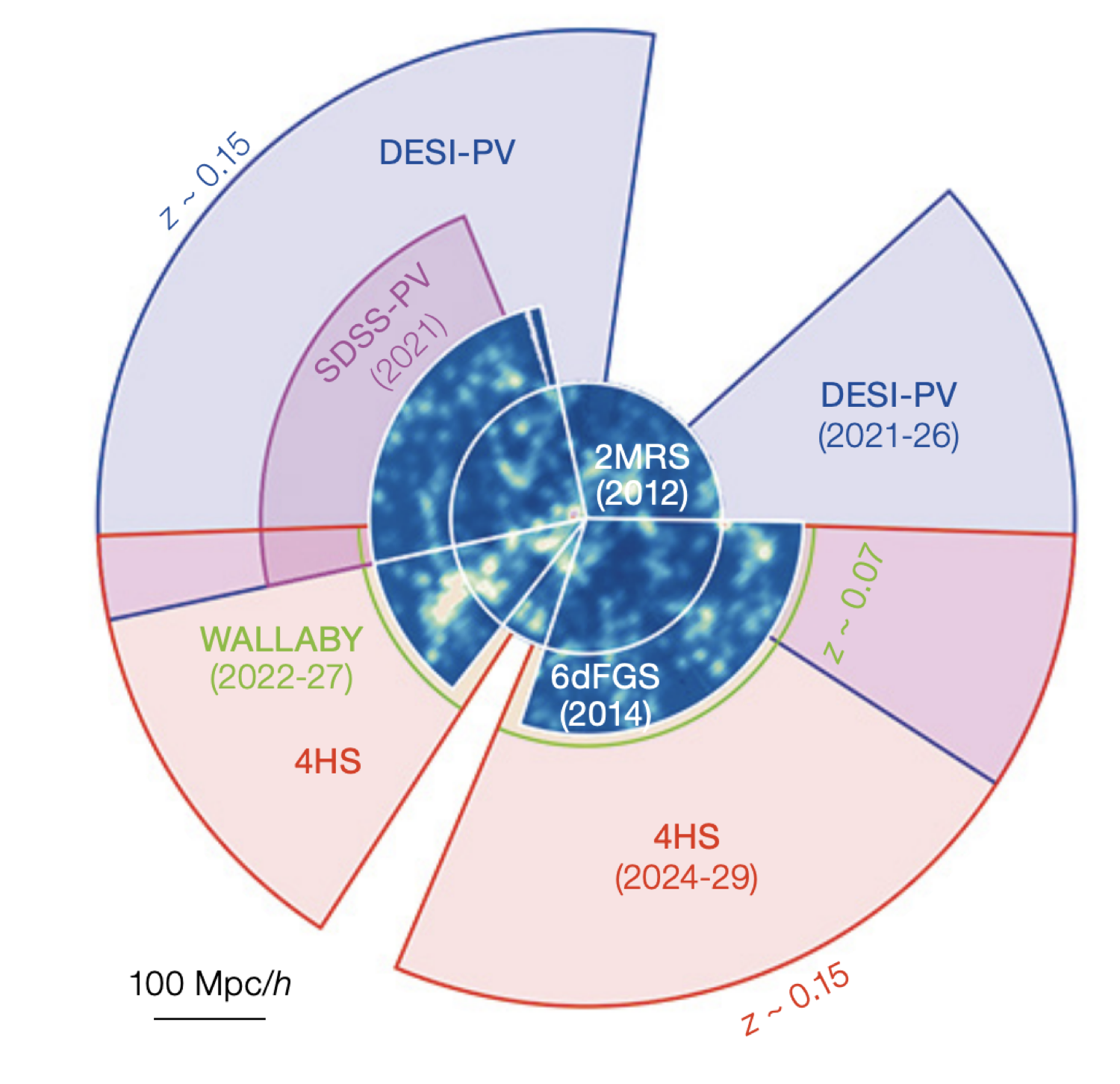}
    \caption{An illustrative depiction of the sky coverage and redshift range of some of the peculiar velocity surveys discussed within this chapter, including 2MTF, 6dFGS, SDSS, DESI, 4HS and WALLABY. Associated years indicate either the expected lifespan of the survey, if the survey is ongoing or yet to begin, or the time of release of the PV dataset if the survey is finished. Credit: Taylor et al. (2023).}
    \label{fig:pv-sky}
\end{figure}

We show how these future peculiar velocity surveys, as well as some of the current surveys, cover the sky, as depicted by a two-dimensional circle, in Fig. \ref{fig:pv-sky}.

The Rubin Observatory Legacy Survey of Space and Time (LSST) and the Zwicky Transient Facility (ZTF) are two large photometric supernovae surveys that will together greatly expand upon current SNe Ia samples, increasing the number of measured Type Ia supernovae to the order of millions. In many ways ZTF is a precursor for LSST. ZTF is more focused on the lower redshift range, complementary with wide-field optical surveys, while LSST probes the larger volumes at higher redshift. The two surveys are located in different hemispheres, LSST in the south and ZTF in the north, and so are complementary with one another, giving us a map of structure in the local universe over the whole sky. ZTF observes with a three-day cadence while LSST will be considerably quicker, being able to image each of its `fields' in the southern sky a minimum of 825 times over its 10-year lifespan. The most relevant SNe Ia for peculiar velocity cosmology are those with redshift $z < 0.3$, where the magnitude of the peculiar velocities are not yet negligible compared to the recession velocity and distance errors are not yet dominant. Currently, SNe Ia are the only reliable method by which we can probe the peculiar velocity field directly beyond redshifts of $z \sim 0.15$ where the scatter in the TF and FP relations render their distance measurements too inaccurate. Current ZTF datasets contain around 3600 supernovae below $z < 0.06$, while LSST is estimated to obtain orders of magnitude more supernovae out to redshift $z < 0.35$.

The data produced by current peculiar velocity surveys, along with that to come from the future surveys described above, will cover almost the entire sky out to redshift $z = 0.3$ with varying levels of completeness. This will be sufficient for us to be able to measure the growth rate of large-scale structure with $\lesssim 3\%$ measurement error. This kind of precision is comparable to that obtained for measurements of the expansion history of the universe.
We will be able to leverage these exquisite measurements of cosmic growth and expansion in order to perform the most accurate tests of the standard cosmological model yet, enabling us to place the most meaningful constraints on any proposed modifications to GR or alternatives to dark energy.

\section{Summary}\label{sec:summary}

In this chapter, we have covered the field of peculiar velocity cosmology, the surveys designed to collect PV data, and the science that is being done with these datasets. We have discussed the mathematical underpinnings of peculiar velocity science and how the velocity field is intimately linked to the underlying density field via the theory of gravitational instability, in so doing explaining the physical processes by which peculiar velocities inform us about the gravitational influence of large-scale structure. We have also covered the various methods of obtaining redshift-independent distances and velocities for galaxies in the local universe, such as galaxy scaling relations like the Tully-Fisher relation and the Fundamental Plane relation, as well as `standard candles' like Type Ia supernovae, and the multitude of surveys that have been dedicated to collating as many of these velocities as possible in the last two decades. We have also discussed the different methodologies invoked by various groups to constrain cosmological parameters using this data, either using the statistical properties of the velocity and density fields themselves to extract information or directly comparing results from one field with predictions made with the other to the same ends. 

In the timeframe covered here, from the SFI++ catalogue being the pinnacle of the field in 2007, to 4HS and DESI measuring 100 times as many velocities over most of the sky out to a redshift of $z \sim 0.15$ by 2030, we have made great leaps in the blink of a cosmic eye. In 1995, \citeauthor{Strauss1995} stated that we were in a ``golden age of exploration'' in reference to the galaxy redshift surveys of the time and their impact on our awareness of the distribution of galaxies and structures in the local universe, but that phrase has surely never been more applicable than it is today. The galaxy redshift surveys of today and their associated peculiar velocity surveys will enable us to map out the local velocity and density fields with precision that would have been unthinkable in 1995, and we will soon be able to measure the growth rate of large-scale structure with the same accuracy as we do the expansion history of the universe. We have never been in a better position to confirm or rule out the existence of missing physics or entirely new theories of gravity, to improve the cosmological constraints made with gravitational waves and SNe Ia, and to understand our place in the flow of the local universe.

\begin{ack}[Acknowledgements]
RRJT would like to acknowledge support received through Australian Research Council Discovery Project DP220101610.
\end{ack}

\def\aj{AJ}                   
\def\araa{ARA\&A}             
\def\apj{ApJ}                 
\def\apjl{ApJ}                
\def\apjs{ApJS}               
\def\ao{Appl.Optics}          
\def\apss{Ap\&SS}             
\def\aap{A\&A}                
\def\aapr{A\&A~Rev.}          
\def\aaps{A\&AS}              
\def\azh{AZh}                 
\def\baas{BAAS}
\def\jcap{JCAP}
\def\jrasc{JRASC}             
\def\memras{MmRAS}
\def\na{New Astronomy}
\def\nat{Nature}
\def\mnras{MNRAS}             
\def\pra{Phys.Rev.A}          
\def\prb{Phys.Rev.B}          
\def\prc{Phys.Rev.C}          
\def\prd{Phys.Rev.D}          
\def\prl{Phys.Rev.Lett}
\def\pasa{PASA}
\def\pasp{PASP}               
\def\pasj{PASJ}
\def\physrep{Phys. Repts.}
\def\qjras{QJRAS}             
\def\skytel{S\&T}             
\def\solphys{Solar~Phys.}     
\def\sovast{Soviet~Ast.}      
\def\ssr{Space~Sci.Rev.}      
\def\zap{ZAp}                 
\let\astap=\aap
\let\apjlett=\apjl
\let\apjsupp=\apjs

\bibliographystyle{Harvard}
\bibliography{reference}

\section{Further Information}
\subsection{Figure credits}
The figures in this document that were taken from anther publication are listed here
\begin{itemize}
    \item Fig. 1 -- \cite{Huterer2023}
    \item Fig. 2 -- \cite{Saulder2023}
    \item Fig. 3 -- \cite{Hoffman2024}
    \item Fig. 4 -- \cite{CourtoisTF}
    \item Fig. 5 -- \cite{Said2020}
    \item Fig. 7 -- \cite{Courtois2023}
    \item Fig. 9 -- \cite{Taylor2023}
\end{itemize}
\subsection{General info.}
For more information about general cosmology or astrophysical concepts, see:
\begin{itemize}
    \item \cite{Leavitt1912} -- period-luminosity relation for Cepheids
    \item \cite{Einstein1916} -- General relativity
    \item \cite{Guth1981} -- seminal Inflationary theory paper
    \item \cite{Riess1998, Perlmutter1999} -- accelerated rate of expansion 
    \item \cite{Carroll2001} -- cosmological constant
    \item \cite{Kepler2007} -- white dwarfs
    \item \cite{PlanckCollab2018} -- Recent cosmology results from Planck satellite 
\end{itemize}
\subsection{Modified gravity}
For more information about modified gravity, see
\begin{itemize}
    \item \cite{Dvali2000} -- DGP gravity
    \item \cite{Linder2007} -- $\gamma$ parameterisation of gravity
    \item \cite{Mirzatuny2019} -- f(R) gravity
    \item \cite{Baker2014} -- scale-dependent growth rate
    \item \cite{Khoury2004, Will2014} -- screening mechanisms
\end{itemize}
\subsection{Surveys and research groups}
For more inforamtion regarding the surveys discussed in this document, see
\begin{itemize}
    \item DESI -- \cite{RuizMacias2020, DESICollab2022, Hahn2023, Saulder2023}
    \item Pantheon+ -- \cite{Scolnic2022}
    \item SH0ES -- \cite{Riess2016, Riess2022}
    \item Union3 -- \cite{Rubin2023}
    \item DES -- \cite{DES2005, DES2024}
    \item Early TF and FP datasets -- \cite{Aaronson1982, Burstein1987, Lucey1988, Faber1989, Mould1991, Willick1991, Mathewson1992}
    \item Mark III -- \cite{Willick1997}
    \item SFI -- \cite{Giovanelli1994, Giovanelli1995, Giovanelli1997, Dale1999}, SFI++ -- \cite{Springob2007, Springob2009}
    \item 6dFGS -- \cite{Jones2004, Jones2005, Jones2009, Springob2014}
    \item 2MASS -- \cite{Skrutskie2006}, 2MTF -- \cite{Hong2019}, 2MRS -- \cite{Huchra2012}
    \item ALFALFA -- \cite{Giovanelli2005, Haynes2018}
    \item SDSS -- \cite{Howlett2022}
    \item Cosmicflows -- \cite{Tully2016, Tully2022}
    \item ZTF -- \cite{Bellm2019}
    \item LSST -- \cite{Ivezic2019}
    \item 4HS -- \cite{Taylor2023}
    \item WALLABY -- \cite{Koribalski2020}
\end{itemize}
\subsection{Standard candles and sirens}
For more information regarding standard candles and standard sirens, see
\begin{itemize}
    \item \cite{Schutz1986} -- gravitational waves as standard sirens
    \item \cite{Branch1992, Phillips1993} -- type Ia supernovae as standard candles
    \item \cite{Hjorth2017, Howlett2020, Nicolaou2020} -- peculiar velocity of host galaxy of GW170187
    \item \cite{Hui2006, TDavis2011} -- effects of peculiar velocities on Hubble diagram
    \item \cite{Carr2022, Peterson2022} -- peculiar velocities in Pantheon+
    \item \cite{Abell2009, Garcia2020} -- LSST science
    \item \cite{Rigault2024} -- ZTF data
\end{itemize}
\subsection{Other distance indicators}
For more information regarding other distance indicators, see
\begin{itemize}
    \item \cite{Freedman2019} -- tip of the red giant branch
    \item \cite{Tonry1988} -- surface brightness fluctuations
    \item \cite{Madore2020} -- J-region asymptotic giant branch
    \item \cite{Ripoche2020} -- carbon stars
    \item \cite{Gao2016} -- megamasers
    \item \cite{Rejkuba2012} -- globular cluster luminosity function
\end{itemize}
\subsection{Velocity and density fields}
For more information regarding the velocity and density fields, see:
\begin{itemize}
    \item \cite{McDonald2009, Blake2013, Koda2014} -- reduced sample variance in two-field analyses
    \item \cite{Watkins2009, Feldman2010} -- differing wavenumber sensitivity of velocity and density fields
    \item \cite{Dekel1999, Springob2014, Tully2014, Graziani2019} -- cosmography
    \item \cite{Park2006, Okumura2014} -- momentum power spectrum
\end{itemize}
\subsection{Bulk flow}
For more information regarding bulk flow measurements, see
\begin{itemize}
    \item \cite{Hong2014, Scrimgeour2016, Qin2021} -- agreement with $\Lambda$CDM
    \item \cite{Watkins2009, Feldman2010, Whitford2023} -- disagreement with $\Lambda$CDM
\end{itemize}
\subsection{Reconstructions}
For more information regarding reconstructions, see
\begin{itemize}
    \item \cite{Nusser1994}
    \item \cite{Hudson1995}
    \item \cite{Zaroubi1999}
    \item \cite{Pike2005}
    \item \cite{Hudson2012}
    \item \cite{Jasche2013}
    \item \cite{Lavaux2016}
    \item \cite{Valade2022, ValadeHamlet} 
    \item \cite{Qin2023}
\end{itemize}
\subsection{Recent results}
For more information regarding the recent results shown in Fig. \ref{fig:fs8-comp}, please see Table \ref{tab:fs8-results}.

\begin{table}[]
    \centering
    \begin{tabular}{c|c|c|c}
        Author(s) & $f\sigma_8$ & Method & $z_{\mathrm{eff}}$\\
        \hline
        \cite{Johnson2014} & 0.418 $\pm$ 0.055 & Maximum-likelihood & 0.000 \\
        \cite{Howlett2MTF} & 0.510$^{+0.090}_{-0.080}$ & Power spectrum & 0.000 \\
        \cite{Lilow2021} & 0.367 $\pm$ 0.060 & Reconstruction & 0.013 \\
        \cite{Boubel2024} & 0.360 $\pm$ 0.030 & Reconstruction & 0.017 \\
        \cite{Davis2011} & 0.310 $\pm$ 0.040 & Reconstruction & 0.020 \\
        \cite{Carrick2015} & 0.401 $\pm$ 0.024 & Reconstruction & 0.020 \\
        \cite{Huterer2017} & 0.428$^{+0.048}_{-0.045}$ & Maximum-likelihood & 0.020 \\
        \cite{Hollinger2024} & 0.362 $\pm$ 0.023 & Reconstruction & 0.020 \\
        \cite{Boruah2020} & 0.400 $\pm$ 0.017 & Reconstruction & 0.022 \\
        \cite{Nusser2017} & 0.400 $\pm$ 0.080 & Correlation function & 0.027 \\
        \cite{Qin2019} & 0.404$^{+0.082}_{-0.081}$ & Correlation function & 0.028 \\
        \cite{Dupuy2019} & 0.430 $\pm$ 0.140 & Correlation function & 0.028 \\
        \cite{Said2020} & 0.311 $\pm$ 0.027 & Reconstruction & 0.035 \\
        \cite{Adams2020} & 0.384 $\pm$ 0.052 & Maximum-likelihood & 0.045 \\
        \cite{Turner2023} & 0.325 $\pm$ 0.075 & Correlation function & 0.045 \\
        \cite{Courtois2023} & 0.380 $\pm$ 0.040 & Correlation function & 0.051 \\
        \cite{Achitouv2017} & 0.420 $\pm$ 0.060 & Correlation function & 0.052 \\
        \cite{Blake2018} & 0.380 $\pm$ 0.120 & Power spectrum & 0.060 \\
        \cite{Beutler2012} & 0.428$^{+0.079}_{-0.068}$ & Redshift-space distortion & 0.067 \\
        \cite{Lai2023} & 0.405$^{+0.076}_{-0.071}$ & Maximum-likelihood & 0.073 \\ 
        \cite{Lyall2024} & 0.329$^{+0.081}_{-0.083}$ & Correlation function & 0.073 \\
    \end{tabular}
    \caption{The measurements of $f\sigma_8$ that are depicted visually in Fig. \ref{fig:fs8-comp} and their associated publications. ordered by the effective redshift of the growth rate measurements either as reported by the authors or surmised from the datasets utilised. Any differences between the reported $z_{\mathrm{eff}}$ and the placement of the measurements in Fig. \ref{fig:fs8-comp} is solely for visual clarity, and the general methodology used by the authors is also included for ease of comparison between the figure and table.}
    \label{tab:fs8-results}
\end{table}
\end{document}